\documentclass[fleqn,usenatbib]{mnras}
\usepackage[T1]{fontenc}
\usepackage{amssymb,epsfig}
\usepackage[]{empheq}
\usepackage{natbib,ifthen}
\usepackage{mathrsfs}
\usepackage{color}
\usepackage{bbold}
\usepackage{csvsimple}
\usepackage{hyperref}       % hyperlinks
\usepackage{url}            % simple URL typesetting
\usepackage{booktabs}       % professional-quality tables
\usepackage{amsfonts}       % blackboard math symbols
\usepackage{amsmath}
\usepackage{graphicx}
\usepackage{wrapfig}
\usepackage{braket}
\usepackage{physics}
\usepackage{algorithm2e}
\usepackage{makecell}

\usepackage{xcolor}

\title{Exoplanet transit search at the detection limit: detection and false alarm vetting pipeline}

\author[J. Robnik, et. al.]{
Jakob Robnik,$^{1}$\thanks{E-mail: jakob\_robnik@berkeley.edu}
Uro\v{s} Seljak$^{1, 2}$,
Jon M. Jenkins$^3$ and
Steve Bryson$^3$
\\
$^{1}$Department of Physics,
University of California, Berkeley, CA 94720, USA\\
$^{2}$Lawrence Berkeley National Laboratory, 1 Cyclotron Road, Berkeley, CA
93720, USA \\
$^{3}$NASA Ames Research Center, Moffett Field, CA 94035, USA
}

\pubyear{2026}

\begin{document}
\label{firstpage}
\pagerange{\pageref{firstpage}--\pageref{lastpage}}
\maketitle

\begin{abstract}
    One of the primary mission goals of the Kepler space telescope was to detect Earth-like terrestrial planets in the habitable zone around Sun-like stars. These planets are at the detection limit, where the Kepler detection and vetting pipeline produced unreliable planet candidates. We present a novel pipeline that improves the removal of localized defects prior to the planet search, improves vetting at the level of individual transits and introduces a Bayes factor test statistic and an algorithm for extracting multiple candidates from a single detection run. We show with injections in the Kepler data that the introduced novelties improve pipeline's completeness at a fixed false alarm rate. We apply the pipeline to the stars with previously identified planet candidates and show that our pipeline successfully recovers the previously confirmed candidates, but flags a considerable portion of unconfirmed candidates as likely false alarms, especially in the long period, low signal-to-noise ratio regime. In particular, several known Earth-like candidates in the habitable zone, such as KOI 8063.01, 8107.01 and 8242.01, are identified as false alarms, which could have a significant impact on the estimates of $\eta_{\oplus}$, i.e., the occurrence of Earth-like planets in the habitable zone.
\end{abstract}

\begin{keywords}
planets and satellites: detection -- methods: statistical -- methods: data analysis
\end{keywords}

\section{Introduction}

\begin{figure*}
    \centering
    \includegraphics[width=\linewidth]{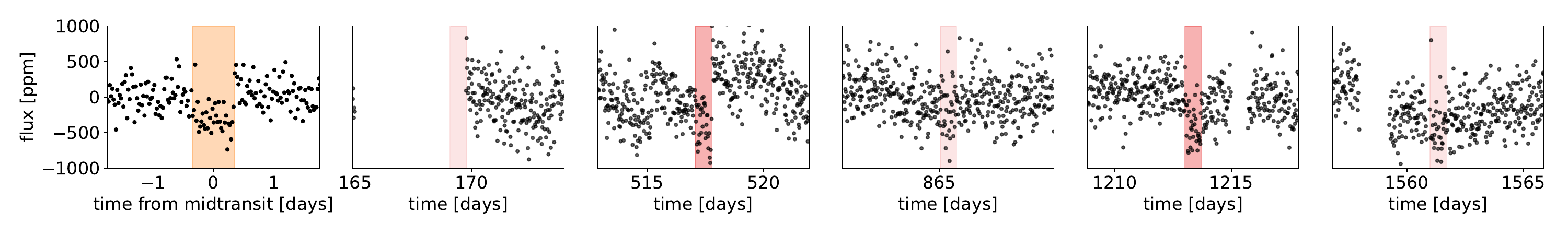}
    \caption{Light curve around the transits of the exoplanet candidate KOI 7894.01 with a Robovetter disposition score of $84 \%$ and reliability of $89\%$ \citep{bryson_occurrence_2021}. Shaded area is the transit region. The left-most panel is the folded light curve, which seems like a genuine exoplanet signature. The next five panels are individual transits of the candidate. The second and the fourth transits are apparently discontinuities in the light curve, which clearly reveals that KOI 7894.01 is a false alarm. The moral of this example is that the individual transits contain important information that is lost in the folded light curve. The method proposed in this work easily discards this candidate, while the existing methods fail.}
    \label{fig: KOI 7894}
\end{figure*}

The Kepler space telescope \citep{borucki_kepler_2010} provided high-quality photometric data with almost continuous four-year coverage for over 150,000 stars, leading to confirmation of over 2000 exoplanets, thus revolutionizing the field \citep{batalha_exploring_2014, winn_occurrence_2015}. However, Kepler fell short of achieving one of its primary mission goals: to robustly characterize the occurrence of Earth-like habitable exoplanets \citep{bryson_why_2025}. Such planets are difficult to detect due to their relatively large periods and small radius, resulting in a faint signal signature. While Kepler's instrumental photometric precision is more than sufficient for detecting such planets \citep{koch_kepler_2010, christiansen_derivation_2012, van_cleve_kepler_2016}, the difficulty is in stellar inherent variability being $\sim 2 \times$ larger than expected at the mission design time \citep{gilliland_kepler_2011, gilliland_kepler_2015} and in the various systematic noise features that can combine to produce planet-like false alarms \citep{thompson_planetary_2018}.

Typical noise systematics are localized glitches in the data, for example, sudden pixel sensitivity drops (SPSDs) or a single measurement outliers \citep{burke_re-evaluating_2019, thompson_planetary_2018}.
Vetting tools like Autovetter \citep{mccauliff_automatic_2015} and the original Exominer \citep{valizadegan_exominer_2022} have difficulty distinguishing such false alarms from the genuine planet transits, because they operate at the level of a folded light curve or at the level of summary statistics, where most of the information about the shape of the individual transits is lost. We note that some of the recent machine learning classifiers are starting to incorporate the information about the individual transits \citep{valizadegan_exominer_2025, tey_identifying_2023}.
The information loss from folding is demonstrated in Figure \ref{fig: KOI 7894}, where both individual transits and a folded light curve are shown for the planet candidate KOI 7894.01. The folded light curve is hard to distinguish from the noisy planet transit, while examining individual transits clearly reveals that 2nd and 4th transits are sudden increase and decrease in the data respectively. When combined they look like a dip, characteristic of an exoplanet transit. 
In an attempt to resolve this issue, a test proposed by \citet{mullally_identifying_2016} was included in the Robovetter \citep{thompson_planetary_2018, coughlin_description_2017}. A hypothesis testing of the exoplanet transit hypothesis against the systematics for \textit{each individual transit} is performed and the event is flagged as spurious if the number of transits for which the systematics hypothesis is not favored is fewer than three. While this is a valuable test, it is still often fails, as can again be seen from Figure \ref{fig: KOI 7894}. KOI 7894.01 has five transits, two of which would have been identified as likely false alarms by the method proposed in \citet{mullally_identifying_2016}, but the number of the remaining transits is still three, so the event would not have been flagged.

{%\color{red}
A different, non-parametric, approach was taken in \citet{ivashtenko_independent_2025}, where first the residuals and the shape of all transit events are tested to identify spurious events which are then omitted from the further analysis. Further, the distribution of the a single-transit Signal-to-Noise Ratio (SNR) is determined empirically and then a transformation is applied to all the single-transit SNRs which maps the empirical distribution to a standard Gaussian distribution.} Effectively this reduces the impact of individual transits that would contribute strongly to the overall SNR, regardless of whether these transits are spurious or not. \citet{ivashtenko_independent_2025} show that while this is better than ignoring the issue altogether, it is also harmful for the completeness of the planet search and reduces the probability of detecting long-period planets (with only three transits) below 10\% for high SNR events and practically to 0 for low SNR.

In this work we propose to examine individual transits, determine the spurious transits and flag the event if the \textit{reduction in SNR} after removing those transits is too large. This is a better strategy than counting the transits because it also takes into account how much these spurious transits contribute to the overall significance. On the example of Figure \ref{fig: KOI 7894} this test clearly shows that the two spurious transits constitute a majority of the total SNR and results in KOI 7894.01 being flagged as a likely false alarm.
%Transits will be considered spurious if they are matched better with false alarm templates than with the planet template or if they occur very close to a large gap in the data, these regions being known for glitches. 

An obstacle to the wide accessibility of the Kepler pipeline is its computational cost. The dominant contribution is the detection algorithm that folds the light curve at a densely spaced grid in planet's period and phase. To find multiple transit events, the already found events are masked out and the detection algorithm reapplied \citep{Jenkins2020}.
For planets close to the detection limit this is expensive, because the true planets are hidden among other false events, each of which costs one run of the detection algorithm. In this work we propose an alternative, which uses a single run of the detection algorithm to create a list of potential transit events and then iteratively removes the less significant events, if they are caused by the more significant ones. 

Combining these innovations with an already established elements from \citet{Jenkins2020, ivashtenko_independent_2025} we construct a detection and false alarm vetting pipeline. We formulate the pipeline in three parts: preprocessing, detection and vetting in Sections \ref{sec: preprocessing}, \ref{sec: detection} and \ref{sec: vetting} respectively.
In Section \ref{sec: ROC} we show that the proposed additions to the pipeline improve completeness at a fixed false alarm rate by performing injections in the Kepler data. In Section \ref{sec: rean} we apply the pipeline to the Kepler stars with previously detected planet candidates and show that it independently identifies the confirmed planets, but eliminates a considerable portion of the unconfirmed planets as likely false alarms.
The resulting pipeline is very fast. Our python implementation with a single CPU on a single star takes 1 minute for preprocessing, 3.5 minutes for detection and 2.5 seconds for false alarm vetting per TCE. 
{%\color{red}
The implementation is publicly accessible at \url{https://github.com/JakobRobnik/exo-probability.git}}

\section{Modeling systematics} \label{sec: setup}

\begin{figure}
    \centering
    \includegraphics[width=\linewidth]{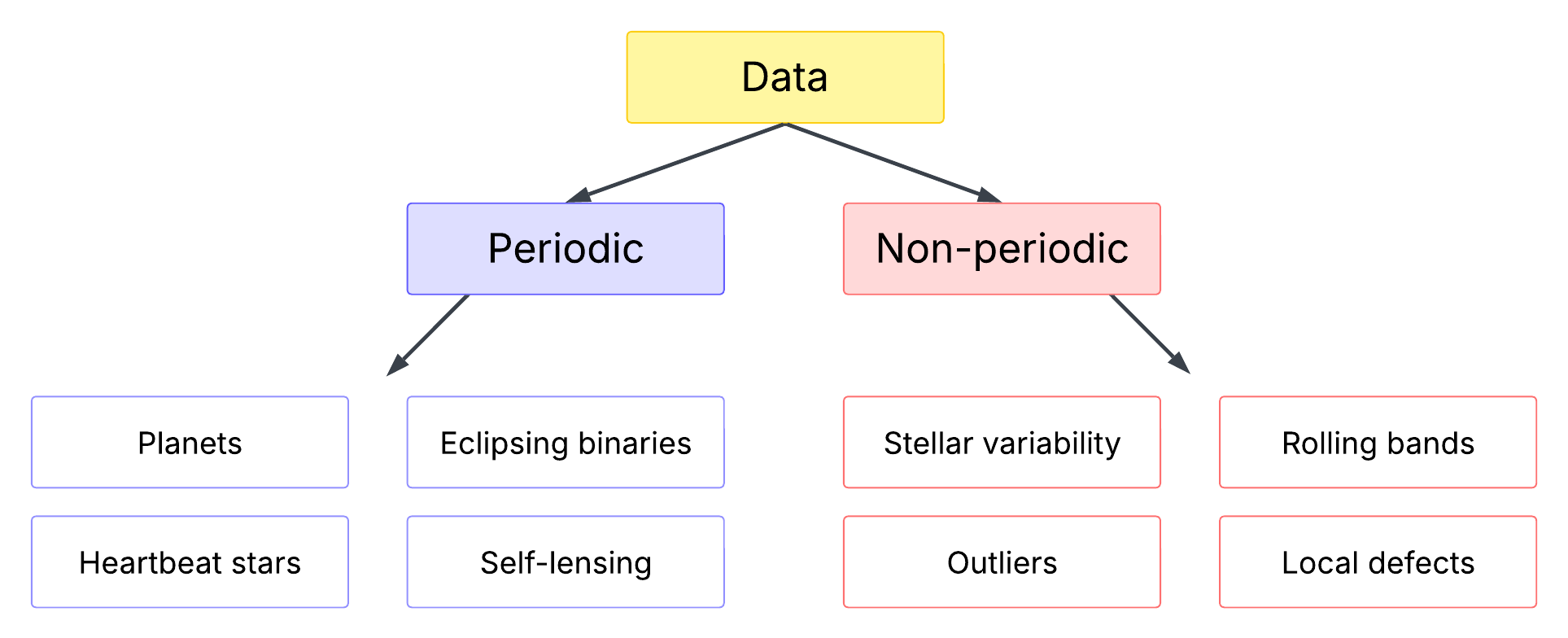}
    \caption{Data is composed of non-periodic and possibly of  periodic sources. Periodic sources are mostly of astrophysical origin and are distinguished from the planets by the Bayesian hypothesis testing. Non-periodic sources are mostly noise systematics and are less well understood. They need to be filtered out, which is the main focus of this work.}
    \label{fig:classification}
\end{figure}

\begin{figure*}
     \centering
     \includegraphics[width=\linewidth]{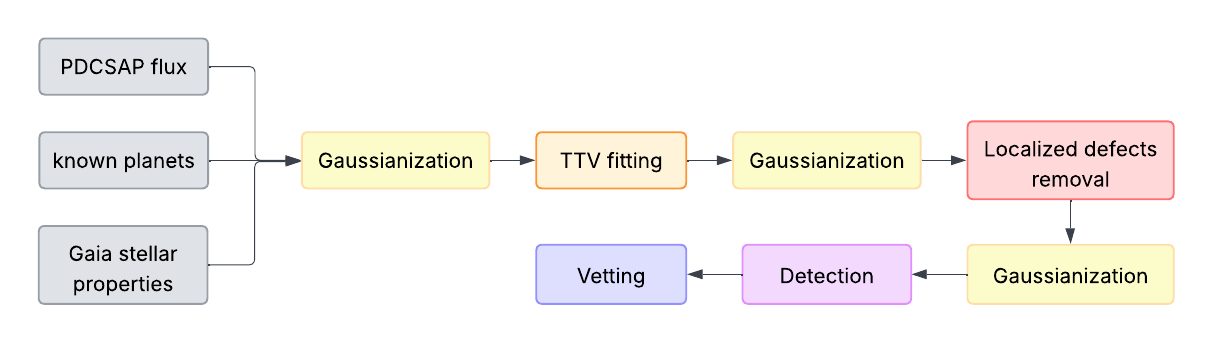}
     \caption{Flowchart of the data analysis. The pipeline is composed of three stages: preprocessing (yellow, orange and red), detection (pink) and vetting (purple).}
     \label{fig:flowchart}
\end{figure*}
%We will be using the PDCSAP flux.
{%%\color{red}
Kepler time-series data is a series of flux measurements $\{ x_i \vert i \in \mathcal{I} \}$ at times $\{ t_i \vert i \in \mathcal{I}\}$, where $\mathcal{I} \subset \mathbb{N}$. Times are equally spaced, meaning that $t_i = t_0 + i \Delta$, where $\Delta$ is $29.42$ minutes for long cadence Kepler data. Ideally, $\mathcal{I} = 1, 2, \ldots N$, but in practice some of the data points are missing.} The flux is composed of non-periodic sources, which are mostly noise systematics and potentially contains periodic sources, which are mostly of astrophysical origin. This is illustrated in Figure \ref{fig:classification}.
Periodic sources can be planet transits, eclipsing binaries, heartbeat stars, self-lensing signal, etc. To differentiate these scenarios from each other Bayesian hypothesis testing is typically used, because the model for these scenarios is relatively well-understood and informative prior and prior odds can be used. Several packages have been developed for this purpose \citep{giacalone_vetting_2021, morton_vespa_2015, torres_testing_2004}.
The focus of this paper is filtering out the non-periodic sources. We start by modeling various non-periodic sources.

%We will decompose 
% \begin{equation}
%     x_i = x_i^{(variability)} + x_i^{(outliers)} + x_i^{(defects)} + x_i^{(periodic)}
% \end{equation}
% and describe the model for the terms above. 

%, which is the preferred and standard method in this setting \citep{giacalone_vetting_2021}. Bayesian hypothesis testing is sensitive to model and misspecification, making it difficult to apply it to the non-periodic noise systematics, however the astrophysical periodic features are physically relatively well understood and their model is reliable. 

% \begin{align} \nonumber
%     \text{data = correlated Gaussian noise + non-Gaussian outliers} \\ \nonumber
%     \text{+ localized defects + signal},
% \end{align}

\textbf{Stellar variability} is modeled as a stationary zero-mean Gaussian noise \citep{jenkins_impact_2002, foreman-mackey_fast_2017, robnik_kepler_2020}.
%and denote it by $x^{(variability)}$.
As such, it is completely characterized by its covariance matrix $\Sigma_{ij} = \langle x_i x_j \rangle$ .
% \begin{equation}
%     x^{(variability)} \sim \mathcal{N}(0, \Sigma_{ij}).
% \end{equation}
Stationarity implies that $\Sigma_{ij}$ depends only on $\vert i-j \vert$, which makes $\Sigma$ diagonal in the Fourier basis. Its diagonal is called the power spectrum $\mathcal{P}$, which we will infer directly from the data for each star separately. 
% The likelihood in the Fourier basis is then
% \begin{equation}
%     - \log p(x) = \frac{1}{2}\sum_{k} \frac{\vert \FT \{y \}(\nu)\vert^2}{\mathcal{P}(\nu)} ,
% \end{equation}
% where $\mathcal{F}$ is the discrete Fourier transform.

% It will be treated by a matched filter, as in \citet{robnik_matched_2021}. 

\textbf{The white non-gaussian noise} component is modeled as a mixture of a Gaussian distribution and an outlier distribution which we model as a non-central t-distribution (NCT). NCT is a distribution with non-symmetric power law tails, capturing the fact that Kepler's negative outliers have different statistical properties than the positive outliers \citep{robnik_matched_2021}. The parameters of the NCT distribution, the Gaussian distribution and the mixture proportion are inferred directly from the data.

\textbf{Localized defects} are the occasional systematics such as sudden-pixel sensitivity drop (SPSD), artefacts close to the large gap edges, etc.
% We will develop an analytic form for these defects and construct a template bank. In the preprocessing stage \ref{sec: preprocessing} we will perform a matched filter to identify the most significant defects, which we will mask out to prevent them from triggering to many false detections in the detection stage. The remaining defects will be attended to in the vetting stage \ref{sec: vetting}.
SPSD has a shape of a sudden drop, followed by a recovery to the original flux level. We model it by an exponential template \citep{mullally_identifying_2016},
\begin{equation}
    s(t \vert A, t_0, \tau) = 
        A \, H(t - t_0) \, e^{- (t - t_0) / \tau},
\end{equation}
where $H(t)$ is the Heaviside function, $A$ is the amplitude of the drop, $t_0$ is its epoch, and $\tau$ the time scale of the recovery.
We also allow for positive amplitude, which corresponds to thermal transients \citep{Jenkins2020} {%\color{red}
and stellar flares \citep{davenport_kepler_2014}.}
%We also allow for negative $\tau$...
We will also consider the stepwise-discontinuity,
\begin{equation}
    s(t \vert A, t_0, \tau) = A \, \mathrm{sign}(t-t_0) \, e^{- \vert t - t_0 \vert / \tau },
\end{equation}
where $A$ can again be negative. These two models are as in \citet{mullally_identifying_2016}, except that \citet{mullally_identifying_2016} originally modeled stellar variability as a local parabola which was later updated to the more principled global model using the Gaussian process \citep{thompson_planetary_2018}. We use the latter approach.

\section{Preprocessing} \label{sec: preprocessing}

The purpose of this stage is to prepare the light curve for the planet detection algorithm by eliminating the impact of non-Gaussian outliers, known planet transits and very significant localized defects.

As input we take for each star:
\begin{itemize}
    
    \item PDCSAP light curve \citep{jenkins_initial_2010, Jenkins2020}, where long term trends have been eliminated.
    We normalize light curve in different quarters as described in \citet{robnik_kepler_2020}, to get an evenly spaced time series, with unit variance of the Gaussian part of the distribution and zero average. 
    
    \item {%\color{red} 
    Properties of the known planet candidates} from \citet{christiansen_nasa_2025}: period, epoch and transit duration. Candidates' transits will be masked out as described in Section \ref{sec: ttv}.
    
    \item Stellar properties: density, radius, effective temperature, metalicity and gravitational acceleration on the surface from \citet{berger_gaia-kepler_2020}. The latter three are used to estimate the coefficients of the quadratic limb darkening model \citep{claret_gravity_2011}. Density is used to construct the transit duration prior, see Section \ref{sec: bayes}. Radius is used to convert the planet transit amplitude to the radius of the planet, see Section \ref{sec: A2r}.
\end{itemize}

\subsection{Outlier Gaussianization}
\label{sec: gaussianization}

To neutralize the impact of the outliers we will process the data $x$ through a transformation $x \mapsto \Psi(x)$ where $\Psi$ is chosen in such a way that 
\begin{enumerate}
    \item $\Psi(x)$ is standard Gaussian distributed if $x$ is locally white (non-Gaussian) noise from the Kepler data
    \item $\Psi$ acts as an identity if $x$ is locally a correlated structure such as planet transits
\end{enumerate} 
In this way the isolated outliers are removed, but the yet unidentified planet transits are unaffected. Such $\Psi$ was designed in \citet{robnik_matched_2021} and is reviewed in Appendix \ref{appendix: gaussianization}. 
To make sure that the outliers are affected even if they are embedded in correlated stellar variability $s$ we first fit the stellar variability, remove it from the data, apply $\Psi$ and then add the stellar variability back:
\begin{equation} \label{eq: gaussianization}
    x \mapsto \Psi(x - s) + s.
\end{equation}
Note that we do not remove stellar variability prior to the planet search because stellar variability fit might also partially fit the yet unidentified planet transits and thus reduce their significance. Instead, stellar variability will be accounted for by the non-white matched filter.
We model stellar variability as a stationary Gaussian noise, which is equivalent to the Fourier Gaussian process (FGP; \citealt{robnik_kepler_2020}). Fitting a Gaussian process requires the power spectrum which we infer directly from the data, as described in Section \ref{sec: PSD estimation}.

The FGP fit to the data can be spoiled in the presence of the outliers and gaps with missing data. Therefore the Equation \eqref{eq: gaussianization} is applied iteratively (3 times): one first fits stellar variability, applies Equation \eqref{eq: gaussianization} to eliminate the impact of outliers and then refits stellar variability. 
After each FGP fit, the gaps where data is missing are filled with the FGP fit as in \citet{robnik_matched_2021}. We note that the gap-filling approach from \citep{ivashtenko_independent_2025} or \citep{autoregression} would also be a good alternatives, but we do not adopt it here, because we need the FGP fit to remove the outliers in any case.

The result of this procedure is the data where the impact of the outliers has been eliminated and the gaps in the data have been filled. This is to be contrasted with the approach taken in \citet{ivashtenko_independent_2025}, where the gaussianization transformation is applied at the level of the data convolved with the planet template. \citet{ivashtenko_independent_2025} does not have a mechanism to preserve the correlated structures such as planet transits in the gaussianziation, so they are equally affected and their significance is reduced, to the extent that the probability of detecting a planet with three transits is below $10 \%$, even for high SNR planets.

\subsection{Power spectrum estimation} \label{sec: PSD estimation}

We estimate the power spectrum directly from the data. First we compute the absolute square of the data in the Fourier domain, $\vert \mathcal{F}(x)(\nu)\vert^2$, and then smooth it by a moving average. We improve the estimate by making the width of moving average bin frequency dependent: at large frequencies the power spectrum is flatter so larger bins can be used. We determine the width of the bin adaptively. The guiding principle is that the width of the bin should be as large as possible (so that the variance of the estimate is as small as possible), but not as large as to support an apparent trend in the bin (which the moving average would smooth out and thus introduce bias). For each bin we therefore compute twice the log-likelihood ratio between the hypothesis that the power spectrum in the bin is a constant plus a Gaussian noise against the hypothesis that it is a linear trend plus a Gaussian noise. We start with bins of size 20 and increase the size until twice the log-likelihood ratio in favor of the linear trend hypothesis becomes larger than one. In this way we determine the bandwidth as a function of frequency on a grid of frequencies, which we then smoothly interpolate to all the frequencies. 

\subsection{Reanalyis of large planets: transit timing variations} \label{sec: ttv}
The potential presence of large planet transits might hide the smaller transits in the detection stage, so we first eliminate the impact of known large planets. The catalog from \citet{christiansen_nasa_2025} provides periods, phases and transit durations for known planet candidates. Our goal here is to mask out the transits of those planets. However, due to the planet-planet interactions, some planets have significant transit-timing variations (TTVs). For example, Kepler 90g has one of the transits more than a day away from where it would be for a periodic orbit. The periodic mask would not remove such transits. To avoid this situation we will systematically search for TTVs and only then apply the mask. The resulting TTV catalog is a useful result in its own right and will be addressed in future work. 

For each planet with disposition score CONFIRMED or CANDIDATE we first determine if individual transits of that planet are significant enough to enable TTV analysis. If this is not the case, we proceed with the periodic template. 
The criterion that we use is that expected $SNR$ of the weakest transit should be above 3.5.

If the individual transits are significant enough, we compute the SNR of the individual transits at the ephemerides for the strictly periodic transits. We identify those, whose SNR is statistically significantly smaller than what would be expected if all transits contributed equally to the overall SNR of the candidate. This signifies that the transit is shifted. If the expected transit location is not covered by a gap we conduct a search for the missing transit. We compute a convolution of the data with the transit template and find the best match in the region $\pm 0.03 P$ around the expected transit location. %$0.03$ was chosen as a generous margin compared to the known TTVs.
We then take the identified TTVs as an initial condition and optimize the TTV model to fine tune the TTVs. 

Finally, to eliminate the impact of known planets we mask the regions of width $1.3 \tau$ around the transit locations, where $\tau$ is the transit duration.

\subsection{Localized defects} \label{sec: individual}

In the vetting stage we will develop tests that flag planet candidates that are likely caused by localized defects such as SPSDs. However, if there are many localized defects in the data they produce a large number of false detections which makes the vetting stage unnecessary computationally expensive and may even obscure genuine planet transits. We therefore mask out the significant localized defects here, in the preprocessing stage. We note that our input PDCSAP flux went through the PDC module \citep{Jenkins2020} which removes the bulk of the more significant localized defects, but we still commonly observe them in the PDCSAP flux, especially at lower SNR.

We build a template bank for localized defects as in \citep{mullally_identifying_2016} and test each transit individually against the null. 
Both for SPSDs and for stepwise discontinuity the template bank consists of 30 logarithmically spaced values of $\tau$ in the range from 45 minutes to 2 days. $SNR_{FA}(t_i)$ is obtained by convolving the data with the templates and taking the best template at each time $t_i$. 
All peaks in $SNR_{FA}(t)$ above 8 are masked out to prevent them from causing false alarm triggers in the planet search. 

False alarms could have biased the FGP fit so we redo the procedure from Section \ref{sec: gaussianization} and fill the newly masked regions with FGP.

\subsection{Harmonics removal}

Some Kepler stars contain sharp peaks in the power spectrum, known as ``harmonics'' \citep{Jenkins2020}. These peaks are too narrow to be well resolved in the power spectrum estimation, which can cause suboptimal matched filtering. We will therefore remove the harmonics, prior to the planet search. In the Kepler pipeline they are fitted out by a phase shifting harmonic function. Here we follow \citet{ivashtenko_independent_2025}. 
We first compute the unsmoothed power spectrum $\vert \mathcal{F}(x)(\nu) \vert^2$, where $\mathcal{F}$ is the discrete Fourier transform. We identify sharp peaks in the unsmoothed power spectrum by a wavelet transform (as implemented in \citet{virtanen_scipy_2020}) and remove peaks with $SNR > 50$. We multiply the smoothened power spectrum by the absolute square of the Notch filter,
\begin{equation}
    \mathcal{P}(\nu) \mapsto \mathcal{P}(\nu) \prod_{k=1}^K \bigg( 1 + \bigg(\frac{\Delta \nu \nu }{\nu_k^2 - \nu^2} \bigg)^2\bigg),
\end{equation}
where $\{ \nu_k \}_{k=1}^K$ are the frequencies of the identified harmonics. Note that this artificially adds a smooth pole in the power spectrum at the frequencies of the harmonics. In the matched filtering this will suppress the data frequencies close to the harmonics frequencies. The parameter $\Delta \nu$ controls the width of the pole, we set it to 0.07, as in \citet{ivashtenko_independent_2025}.

\section{Detection} \label{sec: detection}

\begin{figure*}
    \centering
    \includegraphics[width=\linewidth]{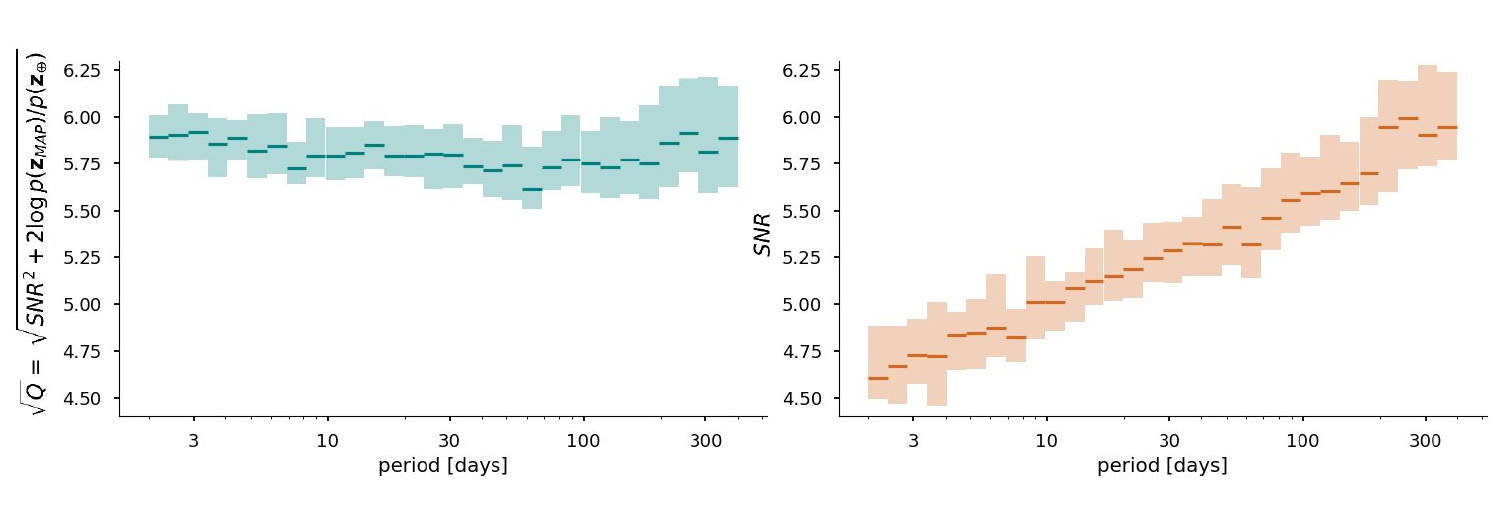}
    \caption{Detection limit as a function of the period of the planet. Noise-only stellar variability simulations are performed and searched for a planet signature. In each period bin, the best candidate is found and its test statistic is shown. The confidence regions are quartiles over 1024 noise simulations, the solid line is the median. Left: Bayes factor is used to construct the test statistic, see Equation \eqref{eq: Q}. The detection limit is roughly independent of the period, so a fixed detection cutoff is close-to-optimal.
    Right: signal-to-noise ratio is used as a test statistic. The detection limit is significantly higher at larger periods. This is because of the look-elsewhere effect \citep{bayer_look-elsewhere_2020, robnik_statistical_2022}: phase range is larger at larger periods, $t_0 < \phi < t_0 + P$, making the prior volume larger and thus increasing the effective multiplicity of trials. 
    This demonstrates that using a fixed detection cutoff on $SNR$ is suboptimal, as it has to be designed to control the false alarms at high periods, while a lower cutoff could be used at lower periods.}
    \label{fig:TestStat}
\end{figure*}

Planet parameters that are well constrained by transit data are planet's period $P$, the phase (epoch) of the first transit $\phi$, transit duration $\tau$ and the amplitude of the transit $A$. Phase is in the range $t_0 < \phi < t_0 + P$.
Test statistics, such as the likelihood ratio (or equivalently the signal-to-noise ratio) are sharply peaked in certain regions of this parameter space. Identifying new planet candidates amounts to identifying those peaks. Peaks that surpass a predefined test statistic threshold are termed threshold crossing events (TCEs; \citealt{Jenkins2020}). Our main novelty here is that we take into account the fact that some of these peaks are correlated, in particular some of the less significant peaks are caused by the more significant peaks, for example at integer multiples of period. We thus make a distinction between the threshold crossing peaks (TCPs), which may be caused by other TCPs and the threshold crossing events (TCEs), which are independent of each other. We first construct a list of TCPs and then convert it to a list of TCEs.
This enables extraction of multiple TCEs from a single scan over planet's parameters, which is significantly cheaper than the approach from \citet{Jenkins2020}, where each TCE is removed from the light curve and the scan is recomputed to identify the next TCE. We note that in the unlikely event that multiple TCEs have an overlapping transit, this transit will contribute to the significance of the most prominent TCE but not to the other TCEs, because it will be removed after the most prominent TCE has been analyzed.

% We note that the Kepler pipeline examines the potential dependencies between the TCEs after the fact by checking if their periods are integer multiples of one another. However, this is not the onle

\subsection{Test statistic}

We have a physically well-motivated prior for the planet parameters. Particularly the transit duration is partially constrained by the Kepler's third law. The prior $p(\boldsymbol{z})$ is derived in Section \ref{sec: prior}.

The optimal test statistic in this kind of hypothesis testing is the Bayes factor (BF; \citealt{robnik_statistical_2022}), because it makes optimal use of the prior information. Bayes factor is the integral of the prior-weighted likelihood ratio over the vicinity of the peak:
\begin{align} \label{eq: bf def}
    B &= \frac{p(x \vert \mathrm{planet})}{p(x \vert \mathrm{no \, planet})}
    = \int \frac{p(x \vert \mathrm{planet \, with \, parameters \, } \boldsymbol{z})}{p(x \vert \mathrm{no \, planet})}p(\boldsymbol{z}) d \boldsymbol{z}.
\end{align}
Computing the Bayes factor with respect to the true null hypothesis is not feasible, because the evidence for the null hypothesis $p(x \vert \mathrm{no \, planet})$ would have to include all the systematic effects, for which an accurate model is not known. We therefore have to resort to the Bayes factor with respect to the null hypothesis which only includes the instrumental noise and stellar variability. The other systematics will then be treated in the vetting stage.
Even so, computing the integral in Equation \eqref{eq: bf def} for every single peak in the parameter space would be too expensive. \citet{ivashtenko_independent_2025} proposed to approximate this integral by a sum over grid points where the likelihood ratio was computed as a side product of the detection algorithm. 
{%\color{red}
Alternatively, \citet{robnik_statistical_2022} showed that for a peak in a parameter space and a parameter space region $Z$, surrounding the peak, the Bayes factor is
\begin{align} \label{eq: Q}
    B(Z) &= V_{\mathrm{post}} e^{Q(Z)/2}, \\ \nonumber
    Q(Z) &= \max_{\boldsymbol{z} \in Z} 2 \log p(\boldsymbol{z}) + SNR^2(\boldsymbol{z}),
\end{align}
where for a given star, $V_{\mathrm{post}}$ is approximately a constant, depending only weakly on the planet parameters.}
Parameters $\boldsymbol{z}$ that maximize the right-hand-side of Equation \eqref{eq: Q}, i.e., the maximum a posteriori (MAP) parameters will be denoted by $\boldsymbol{z}_{MAP}$. Since rescaling a test statistic by a $\boldsymbol{z}$-independent constant of proportionality and applying the logarithm plays no role in the hypothesis testing we will adopt $Q$ instead of $B$ as our test statistic in the detection stage. In vetting stage we will be more accurate and directly compute $B$, since the number of peaks and therefore the cost at that stage will be much lower.

SNR can be obtained by matched filtering the data $x(t)$ with the planet template $S_{\boldsymbol{z}}(t)$:
\begin{equation}
    SNR(\boldsymbol{z}) = \frac
    {\braket
    {x}
    {S_{\boldsymbol{z}}}
    }
    {\sqrt{\braket
    {S_{\boldsymbol{z}}}
    {S_{\boldsymbol{z}}}}.
    }
\end{equation}
Here, the scalar product between the light curves is defined as
\begin{equation} \label{eq: scalar product}
    \braket{x}{y} = \sum_{k} \frac{\mathcal{F}(x)_k^* \mathcal{F}(y)_k}{\mathcal{P}_k},
\end{equation}
where $\mathcal{F}$ is the discrete Fourier transform, $\mathcal{P}$ is the noise power spectrum and $^*$ is the complex conjugate. For more details, see \citet{robnik_matched_2021}. In the literature, a time-dependent power spectrum is often used to account for time varying noise properties \citep{Jenkins2020, ivashtenko_independent_2025}, but we do not observe any gain in performance from using it, see Appendix \ref{sec: nonstat}.

\subsection{Prior} \label{sec: prior}

\paragraph*{Period.} We adopt a uniform period prior from $P_{\mathrm{min}}$ to $P_{\mathrm{max}}$.
We set $P_{\mathrm{max}} = T/2$, where $T$ is the maximal duration of the Kepler data, around 1400 days. Planets with larger periods can safely be ignored, as they cannot produce more than two transits and we require at least three transits for planet confirmation. Here, we set $P_{\mathrm{min}} = 2$ days, as our main interest are longer-period planets, for example the habitable zone planets, but note that the pipeline also works at lower periods.

% We will require at that at least $M = 3$ transits were observed. This changes the phase prior from $\mathcal{U}(0, P)$ 
% to $\mathcal{U}0, \mathrm{min}(P, T - (M-1)P))$ and $P_{\mathrm{max}} = T / (M-1)$. The normalized prior is then
% \begin{equation}
%     p(P, \phi) = \frac{1}{P} \frac{1}{T \log(M / (M-1)) - P_{\mathrm{min}}}
% \end{equation}
\paragraph*{Phase.}
For the phase at a given period we adopt uniform prior from 0 to the period:
\begin{equation}
    p(P, \phi) = \mathcal{U}_{[P_{\mathrm{min}}, P_{\mathrm{max}}]}(P) \, \mathcal{U}_{[t_0, t_0 + P]}(\phi) = \frac{1}{P_{\mathrm{max}} - P_{\mathrm{min}}} \frac{1}{P}
\end{equation}
Note that there is more prior volume at larger periods, because there are more available phases. The false alarm rate with the SNR test statistic is therefore also higher at higher periods. Using SNR as a test statistics is therefore suboptimal, because a single detection threshold does not work well for all the periods. The Bayes factor test statistic automatically takes this into account, as its false alarm rate is independent of the period. We illustrate this effect in Figure \ref{fig:TestStat}, where we simulate noise-only light curves, run detection algorithm developed in this section and show detection limit as a function of period. It can be seen that using SNR as a test statistic makes detection limit strongly dependent on the period, while using $Q$ removes this dependence.

\begin{figure}
    \centering
    \hspace*{-0.3cm}\includegraphics[scale = 0.59]{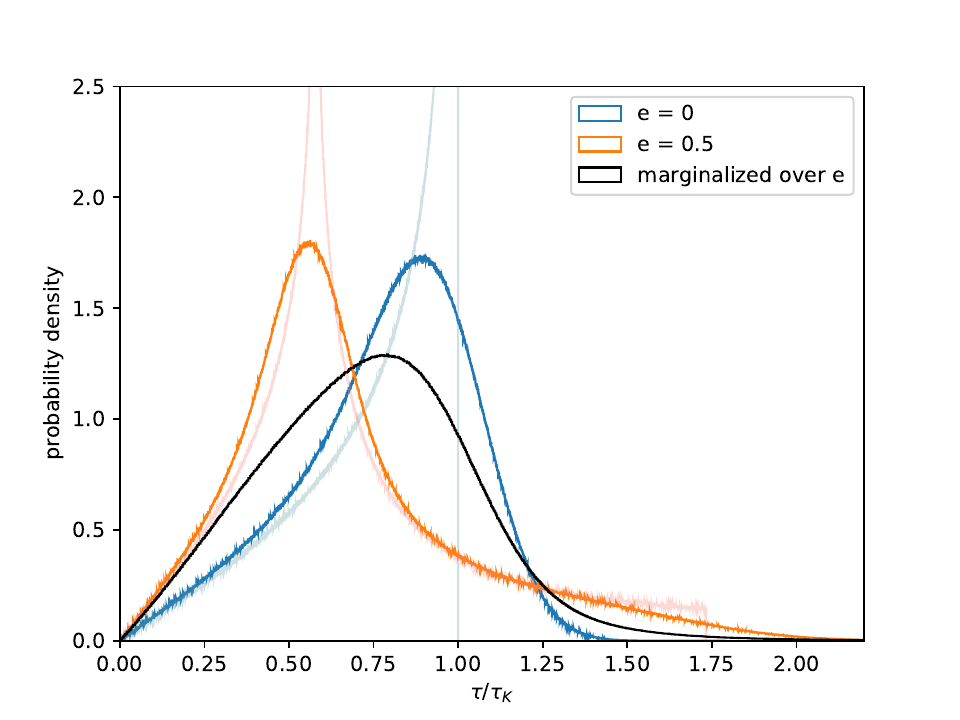}
    \caption{Transit duration prior is shown. Blue and orange lines are showing the prior for planets with orbit eccentricities 0 and 0.5 respectively. Pale versions of both colors show the corresponding priors if $\tau_K$ was known exactly, i.e. if stellar density had no uncertainty. Stronger lines are shown for 15 percent relative error on $\tau_K$. Uncertainty in the $\tau_K$ broadens the prior. The black line is the prior which is marginalized over the eccentricity, assuming a beta distribution prior for the eccentricity with parameters $\alpha = 0.867$, $\beta = 3.03$ as fitted to the Kepler planets \citep{kipping_bayesian_2014}.}
    \label{fig: tau prior}
\end{figure}

\paragraph*{Transit duration.}
The transit duration prior is useful for eliminating false alarms: if the orbit of the planet was circular and perfectly aligned with the line of sight, Kepler's third law would completely fix the transit duration in terms of the planet's period and stellar density $\rho_*$:
\begin{equation} \label{eq: tauK}
    \tau_K = \bigg( \frac{3}{\pi^2 G \rho_*} P \bigg)^{1/3}.
\end{equation}
If the observed transit duration and period are not consistent with Kepler's third law we can eliminate the candidate. In reality, orbits may be inclined and eccentric. Additionally, we typically have around 15 \% relative uncertainty of $\rho_*$ \citep{berger_gaia-kepler_2020} thus $\tau_K$ is not known exactly. This effects make transits durations that are not exactly equal to $\tau_K$ still consistent with Kepler's third law, but nonetheless $\tau$ that deviate significantly from $\tau_K$ are less likely. 
Kepler pipeline applied a transit duration cut based on a similar argument (Sec 9.4.3 of \citet{Jenkins2020}), but it was a binary cut computed at $P_{\mathrm{max}}$ rather than at the observed period. The Bayes factor that we adopt here uses this information optimally in a continuous manner, simply by penalizing transits durations according to the transit duration prior. The prior is shown in Figure \ref{fig: tau prior}, its exact form is derived in Appendix \ref{sec: tau prior}.

\subsection{Identifying TCPs}

We compute $Q$ on a fine grid of periods, phases and transit durations as described in \citet{robnik_matched_2021} and briefly reviewed here. We use a template bank with logarithmically spaced transit durations from 45 min to $1.5 \tau_K(P_{\mathrm{max}})$, see Appendix \ref{appendix: bank}. The data is convolved with each of these templates. On a fine grid of logarithmically spaced trial periods we then 
\begin{itemize}
    \item Fold the convolved light curve, corresponding to the template with transit duration $\tau = \tau_K(P)$. Find the phase which maximizes $Q$. We require the detection to have at least $M=3$ transits, so not all phases need to be computed at periods larger than $T/M$, where $T$ is the time span of the data. It then suffices to only check $t_0 < \phi < t_N +  - (M-1) P$.
    \item On a small grid of phases around the best-fit phase compute $Q$ for all $\tau$ in the template bank. Save the best $Q$ and the corresponding parameters.
\end{itemize}
In this way we obtain best $Q$ at each period. 
We identify a number of highest $Q$-peaks that are above a predefined threshold as TCPs.

\subsection{From TCPs to TCEs}
We convert a list of TCPs to a list of TCEs by an iterative algorithm. The currently most significant peak is a TCE, but the less significant TCPs might be caused by this TCE. The $Q$ values that we computed for those TCPs are therefore only an upper bound for their true test scores $Q$, we will denote them by $\widehat{Q}$. We add the newly identified  TCE to the list of TCEs and mask it from the data. Next we recompute the $Q$ values for the remaining TCPs. We start with the peak with the highest $\widehat{Q}$ and proceed by decreasing $\widehat{Q}$. If at some point we find a peak whose updated $\widehat{Q}$ is higher than all the remaining $\widehat{Q}$ we do not continue updating the $\widehat{Q}$ because we have already identified our next TCE. Also, if any of the peaks' $\widehat{Q}$ falls below the threshold, we remove it from the list. This procedure is shown as a pseudocode in Algorithm \ref{sort}.

\begin{algorithm}
\caption{From TCPs to TCEs. $\mathrm{Mask}(F, \boldsymbol{z})$ is a function that takes a light curve $F$ and masks the transits of a planet with parameters $\boldsymbol{z}$. $Q(F, \boldsymbol{z})$ computes $\widehat{Q}$ at parameters $\boldsymbol{z}$.} \label{sort}
\KwIn{
\begin{itemize}
\item List of TCPs,
$\mathrm{TCP} = \{ (\widehat{Q}^{(k)}, \boldsymbol{z}^{(k)})\}_{k=1}^K$
\item Light curve $F$
\end{itemize}
}
\KwOut{List of TCEs, $\mathrm{TCE} = \{ (Q^{(l)}, \boldsymbol{z}^{(l)} )\}_{l=1}^L$}
\While{$\mathrm{TCP} \neq \{ \}$}{
    Sort $\mathrm{TCP}$ by decreasing $\widehat{Q}$ \;
    $\mathrm{TCE} \gets \mathrm{TCE} \cup \mathrm{TCP}_1$ \;
    $\mathrm{TCP} \gets \mathrm{TCP} \textbackslash \mathrm{TCP}_1$ \;
    $F \gets \text{Mask} (F, \boldsymbol{z}^{(k)})$ \;
    $k \gets 1$ \;
    \Repeat{$\widehat{Q}^{(k)} > \widehat{Q}^{(k+1)}$}{
        $\widehat{Q}^{(k)} \gets Q(F, \boldsymbol{z}^{(k)})$ \;
        $k \gets k + 1$ \;
    }
    $\mathrm{TCP} \gets \mathrm{TCP} \textbackslash \{\mathrm{TCP}_m \vert Q^{(m)} < \text{cutoff} \}$
}
\end{algorithm}

\section{False alarm vetting} \label{sec: vetting}

The number of TCEs is significantly lower than the total number of events, so we can at this point apply techniques that would be to expensive to be applied at the detection stage. First we replace the Bayes factor approximation $Q$ by an accurate calculation of the Bayes factor. Then we will apply additional tests that are designed to eliminate false alarm scenarios that are not included in the Bayes factor evidence. 

\subsection{Bayes factor} \label{sec: bayes}

Bayes factor calculation in the exoplanet context has been proposed by \citet{matesic_gaussian_2024}, using nested sampling \citep{skilling_nested_2004}. A faster alternative is preconditioned Monte Carlo (PocoMC; \citet{karamanis_pocomc_2022}). However, both of these algorithms are general purpose methods, designed to marginalize over a relatively large number of parameters (up to an order of a thousand) and are computationally very expensive, using a few thousand (problem adapted PocoMC) or even millions (nested sampling) likelihood calls. We can take advantage of the fact that the number of exoplanet parameters is very small: integral over the amplitude parameter can be done analytically \citep{robnik_statistical_2022}, leaving us with only three parameters. We adopt Gaussian quadrature integration, which uses only 24 likelihood evaluations, but gives the same accuracy as the sampling approaches \citep{robnik_statistical_2022, robnik_periodicity_2024}.

% \subsection{Geometric probability}

% The probability that a planet is transiting is 
% \begin{equation}
%     h = R_* / a = \pi \tau / P,
% \end{equation}
% where the second step is obtained by noting the similarity with the expresion for the transit duration: $\tau = 2 R_* / v = 2 R_* / (2 \pi a / P)$.
% $h$ is typically around $1\%$. However, all the candidates we considered are in a system where at least one planet has already been confirmed. The planet inclinations are highly correlated, so the fact that one planet is aligned with the line of sight implies a much larger probability that a second planet is also aligned. We will assume the inclination between two planets to be a Rayleigh distribution with the width of $1.5^{\circ}$ \citep{lissauer_architecture_2011,fang_architecture_2012,ballard_kepler_2016}. We compute the probability $P(\mathrm{planet \,2\, observable} \vert \mathrm{planet \,1 \,observable})$ by a Monte Carlo simulation. One could use a faster numerical method \citep{brakensiek_efficient_2016}, but this part of our analysis is not a computational bottleneck. The resulting probability is typically around $40 \%$ so a significant gain compared to the probability of detecting the first planet.
% If more than one planet has already been confirmed, we compute the probability of a new planet by only taking the confirmed planet which is closest to the proposed new planet and again assume the same distribution of mutual inclinations.

\subsection{Examining individual transits}

As demonstrated in Figure \ref{fig: KOI 7894}, folded light curve might look like a planet transit, while individual transits are clearly noise systematics. We will here examine individual transits to eliminate this type of false alarms. 
The strategy is to first identify spurious transits and then compare TCE's significance to the significance that it would have if the spurious transits were removed. 
To identify spurious transits we consider for each transit separately:
\begin{itemize}
    \item \textbf{Localized defects}. We compute the likelihood ratio $l$ between the planet hypothesis and the localized defect hypothesis (as modeled in Section \ref{sec: setup}). For each of the hypotheses we maximize the data convolved with the template bank for that hypothesis over the transit region and over the templates in the bank and thus formulate the likelihood ratio test statistic. We then empirically estimate the p-value using the likelihood ratio of the points outside of the transit as p-value = $\# \{ t \vert l(t) > l,  t \text{ outside the transit}\} / \# \{ t \vert t  \text{ outside the transit}\}$. We consider the transit spurious if this p-value is below $1\%$.
    \item \textbf{Close to large gaps regions} are more likely to contain systematic defects than the other regions due to the effects such as thermally-induced systematics resulting from the changes in the spacecraft orientation. These systematics are largely removed by the preprocessing module \citep{Jenkins2020}, but residuals may persist. We will consider a transit spurious if it is less than 20 hours away from a large gap, where we consider gaps large if they last for at least 20 hours. 
\end{itemize}

For each planet detection we try removing all transits that are considered spurious under any of the above scenarios and flag the object as spurious if this reduces the $SNR^2$ of the object by more than $50 \%$.

\subsection{Transit significance consistency}

\begin{figure}
    \centering
    \includegraphics[width=\linewidth]{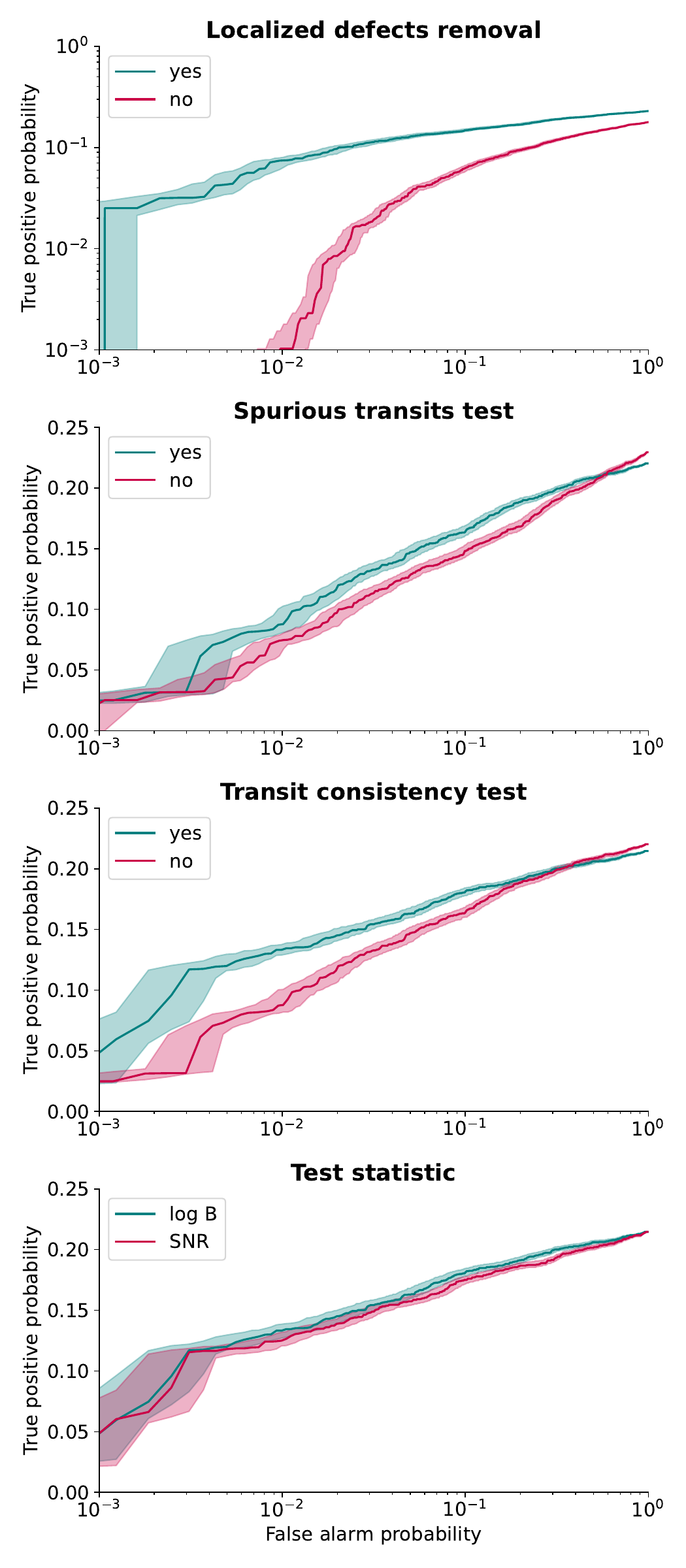}
    \caption{ROC curve showing the true positive probability as a function of false alarm probability on injections in the Kepler data. The uncertainty regions are quartiles obtained by bootstrap.
    Each panel progressively adds the pipeline improvements proposed in this work. 
    First panel: removing localized defects prior to the planet search drastically improves the TPP (by more than an order of magnitude at FAP 0.01).
    Second panel: spurious transits test improves the TPP.
    Third panel: transit significance consistency test improves the TPP.
    Fourth panel: Bayes factor test statistic is only marginally better than SNR. 
    }
    \label{fig:roc}
\end{figure}

As an additional test, we check that individual transits contribute to the evidence the expected amount \citep{seader__2013, Jenkins2020, ivashtenko_independent_2025}. For example if an object of interest has 5 observed transits, but its significance is dominated by a single transit, this is most likely caused by an unmodeled feature in the null hypothesis.

The periodic transits' template $S(t)$ can be broken down in individual transits:
\begin{equation} \label{eq:periodicTemplatedef}
 S(t) = \sum_{n = 1}^{N} S_n(t),
\end{equation}
where $S_n$ is a template of the $n$-th transit, centered at time $\phi + n P$. We will assume that events $S_n$ are well separated, i.e. overlap sums between different events are zero:
\begin{equation}
    \braket{S_n}{S_m} = w_n^2 \delta_{n m},
\end{equation}
where we defined $w_n^2 = \braket{S_n}{S_n}$ and $\braket{\cdot}{\cdot}$ is the scalar product from Equation \eqref{eq: scalar product}. The orthogonality assumption is not trivial due to the noise correlations. The assumption is in principle problematic for planets with short periods, but we verified that even for planets with a 3-day period it would result in only a 3\% bias of the $SNR$. This will suffice for the purpose of this test. Combined $SNR$ using all transits is then
\begin{equation}
    SNR = \frac{\braket{d}{S}}{\sqrt{\braket{S}{S}}} = \frac{\sum_n w_n SNR_n}{\sqrt{\sum_n w_n^2}},
\end{equation}
where $SNR_n$ is a $SNR$ one would get using $n$-th transit only:
\begin{equation}
    SNR_n = \frac{\braket{d}{S_n}}{\sqrt{\braket{S_n}{S_n}}}.
\end{equation}
Note that $w_n$ can be thought of as weights giving the expected relative contribution of each transit towards the detection. For example, if some data is missing during the $n$-th transit, $w_n$ will be smaller. $SNR_n$ is expected to be a realization of a random normal variable with a unit variance and expected value 
\begin{equation}
    E[SNR_n] = \frac{w_n}{\sqrt{\sum_i w_i^2}} SNR.
\end{equation}
We want to test the data against this null hypothesis. We will only test the unit variance expectation, because this is a more powerful statistical test than testing higher moments of the distribution, especially when only a few transits are measured. Let $\sigma^2_{data}$ be the measured variance from the data:
\begin{equation}
    \sigma_{data}^2 = \frac{1}{N} \sum_{n} \big(SNR_n - \mathbb{E}[SNR_n] \big)^2 .
\end{equation}
The null hypothesis is tested by examining the p-value: how likely would $\sigma_{data}^2$ or larger variance arise if the null hypothesis was true. Under the null hypothesis, $N \sigma_{data}^2$ is distributed by a $\chi^2$-distribution with $N$ degrees of freedom. We report the p-value: probability of observing such or larger fluctuations under the planet hypothesis. Low p-value signals a likely false alarm, we use $0.01$ as a threshold.

\begin{figure*}
    \centering
    \includegraphics[width=\linewidth]{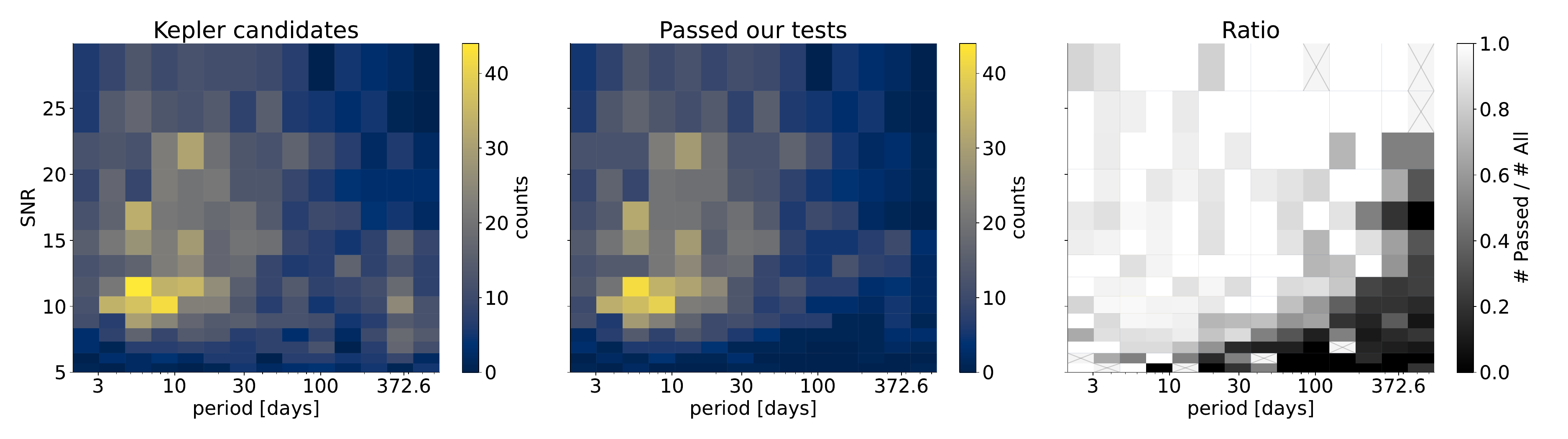}
    \caption{Histogram of all Kepler's planet candidates \citep{christiansen_nasa_2025}, including the unconfirmed ones. Left panel: all Kepler's candidates. An increase in their number is seen at the period of Kepler space telescope's orbit of 372.57 days. This is caused by rolling bands \citep{kolodziejczak_flagging_2010, Jenkins2020}, which are a known source of alarms.
    Middle panel: Kepler's candidates which were independently identified by our pipeline and passed all of our tests. 
    Right panel: fraction of candidates that passed our tests, i.e., the ratio of the other two panels. 
    As can be seen, both pipelines agree on almost all candidates at shorter periods or high SNR. However, at low SNR and long periods many of the Kepler's unconfirmed candidates are not independently identified by our pipeline. This is the region of parameter space where  terrestrial planets in the habitable zone lie.
    }
    \label{fig: candidates}
\end{figure*}

\section{Injection-recovery tests} \label{sec: ROC}

% \begin{figure}
%     \centering
%     \hspace*{-0.3cm}\includegraphics[width=1.05\linewidth]{cand.pdf}
%     \caption{Kepler's unconfirmed planet candidates from \citet{christiansen_nasa_2025} with their radius and period shown. The candidates which are independently identified by our pipeline and passed all of the test are shown in blue, the others in red.}
%     \label{fig: candidates}
% \end{figure}

% \begin{figure*}
%     \centering
% \includegraphics[width=\linewidth]{histConf.pdf}
%     \caption{Histogram of Kepler's confirmed planet candidates \citep{christiansen_nasa_2025} in the period-SNR plane. On the left are all Kepler's confirmed candidates, on the right are those which were independently identified by our pipeline and passed all of our tests. As can be seen, the plots are practically identical: our pipeline and Kepler's pipeline agree on the confirmed planets.}
%     \label{fig: conf}
% \end{figure*}

To test the effectiveness of the proposed false alarm vetting tests we study the Reciever Operating Characterisitc (ROC) with and without those tests. ROC is a plot showing the true positive probability (TPP) as a function of the false alarm probability (FAP). 

We take 5000 randomly chosen Kepler stars and process them through our pipeline. Before the detection stage we invert the light curve (multiply it by -1) to prevent the pipeline from potentially finding new real planets. Using the pipeline we then for each star identify the TCE with the highest test statistic that passed the false alarm vetting stage. 
We do this analysis with and without injecting a planet signature in the data prior to the detection stage of the pipeline. Period of the injected planet is drawn uniformly at random between 200 and 500 days, phase and transit duration from the prior from Section \ref{sec: prior} and the radius of the planet from the normal distribution centered at $2 R_{\oplus}$ with standard deviation $0.2 R_{\oplus}$. We consider the injection successfully recovered if the TCE's period and phase are less than 1\% and 0.5 days from the injected parameters respectively. We determine the TPP, i.e. the fraction of successfully recovered injections, as a function of the test statistic detection threshold. Similarly we obtain the false alarm rate, i.e., the fraction of stars with a TCE in the injectionless run. Matching the TPP with the FAP at the same detection threshold gives the ROC.

ROCs are shown in Figure \ref{fig:roc} for various design choices in the pipeline: Bayes factor vs SNR test statistic, with and without the transit significance consistency test, with and without the spurious transits test and with and without removing the localized defects prior to the detection stage.
Applying the tests and adopting the Bayes factor test statistic improves the probability of detecting the true signal at a fixed FAP. At FAP of 1\% the spurious transits test improves the probability of detecting the true signal by 17\% and the significance consistency test by an additional 52\%. Improvements due to the Bayes factor are small, 7 \%. The largest impact comes from the localized defects removal: a factor of 70.

\section{Results: reevaluation of known planet candidates} \label{sec: rean}

\begin{figure}
    \centering
    \includegraphics[width=\linewidth]{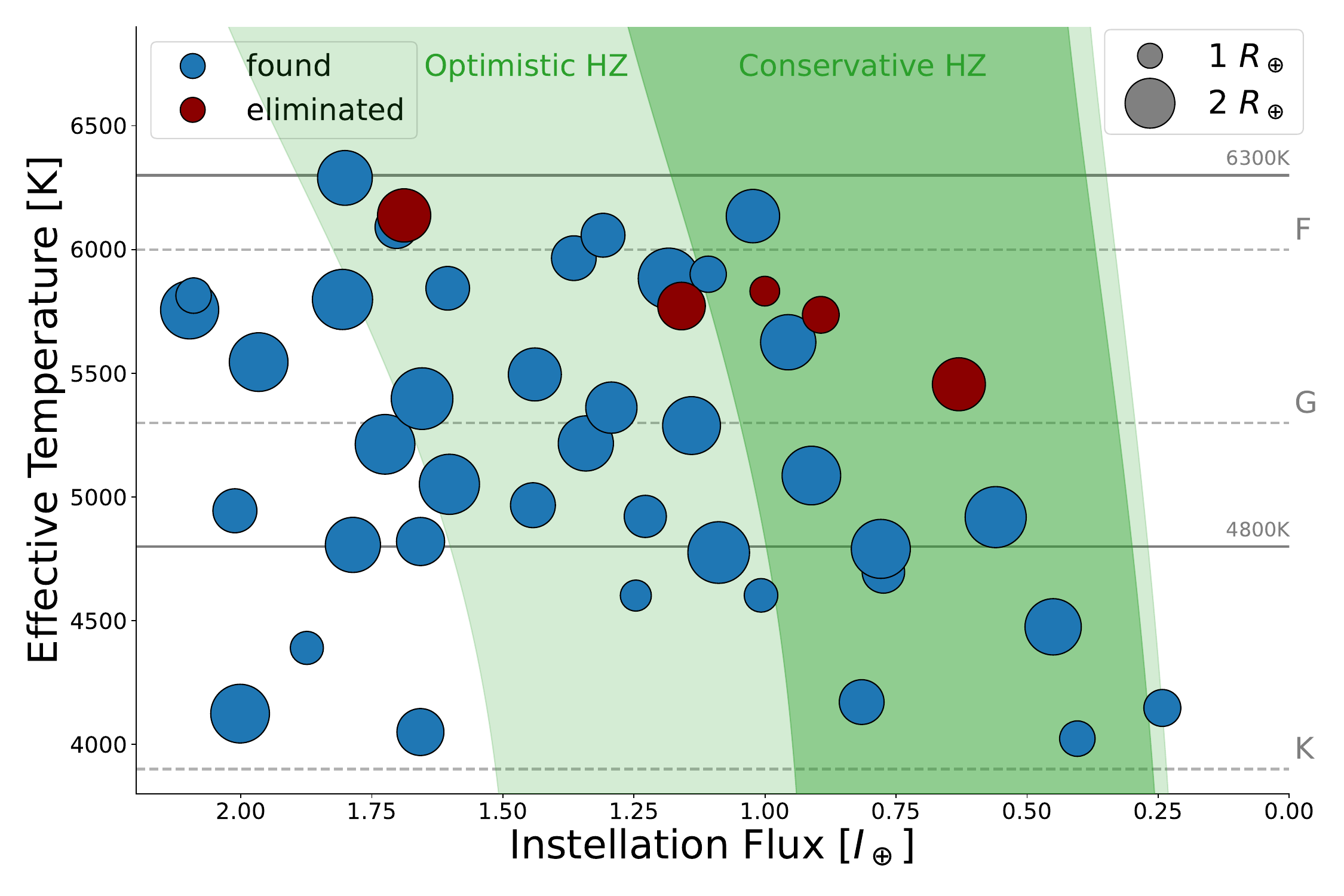}
    \caption{
    Earth-like and super-Earth candidates from \citet{bryson_occurrence_2021} are displayed in the effective temperature–instellation flux plane, as in \citet{bryson_occurrence_2021}. The conservative and optimistic habitable zones defined by \citet{bryson_occurrence_2021} are shown in green. Marker sizes are proportional to the planetary radius. The candidates that were independently identified by our pipeline are shown in blue, the other with red. Out of five eliminated candidates three are in the conservative HZ, one is on the edge between the optimistic and conservative HZ and one in the optimistic HZ. Due to the overall small number of candidates in the habitable zone with radius $< 1.8 R_{\oplus}$ this reduction could have a significant impact on $\eta_{\oplus}$ estimates. 
    }
    \label{fig: hz}
\end{figure}

\begin{figure}
    \centering
    \includegraphics[width=\linewidth]{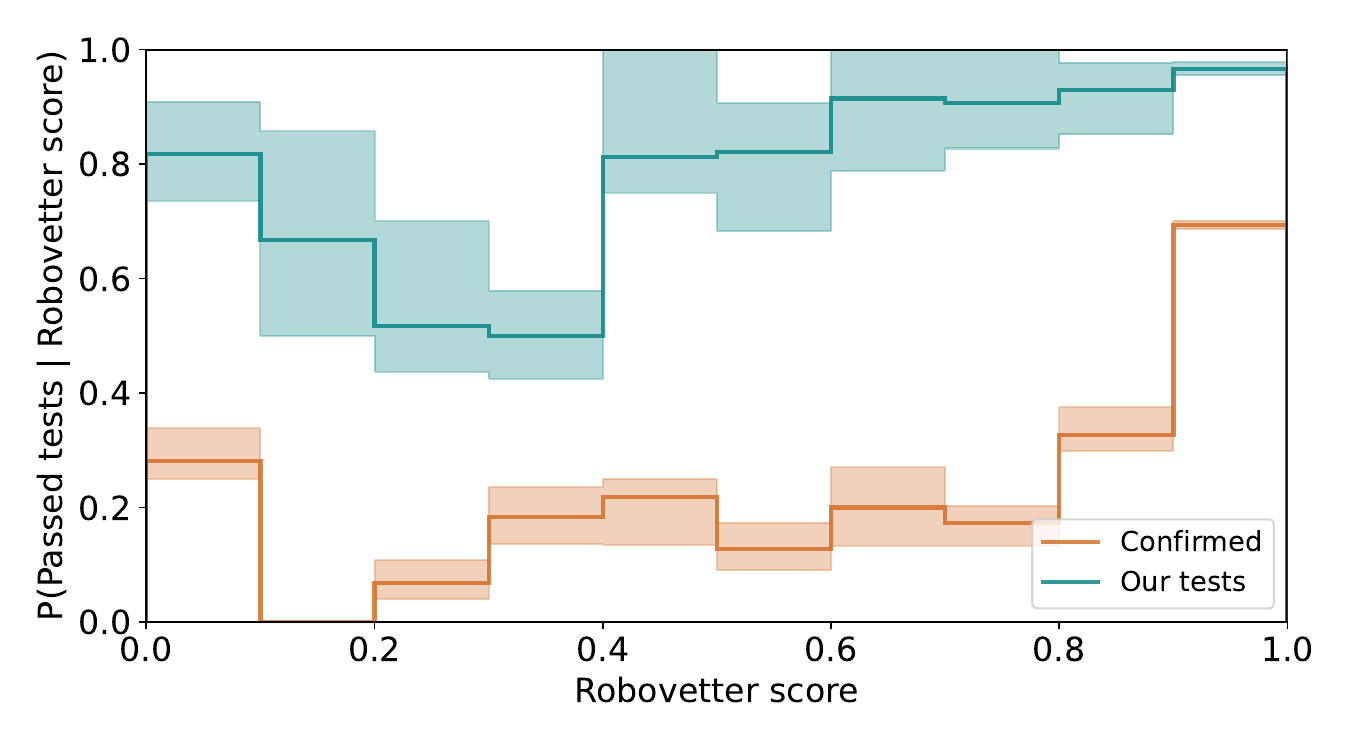}
    \caption{Probability that a given planet candidate is independently identified by our pipeline (teal) or considered a confirmed Kepler planet (orange), given its Robovetter score. The confidence regions are the quartiles obtained by bootstrap. Candidates with any score, including the highest scores,  may fail to trigger our algorithm.}
    \label{fig:robovetter}
\end{figure}

\newcommand{\DoFigure}[2]{%
  \begin{figure*}
    \centering
    \includegraphics[scale=0.3]{KOI#1.pdf}
    \caption{#2}
    \label{fig: KOI#1}
  \end{figure*}
}

\DoFigure{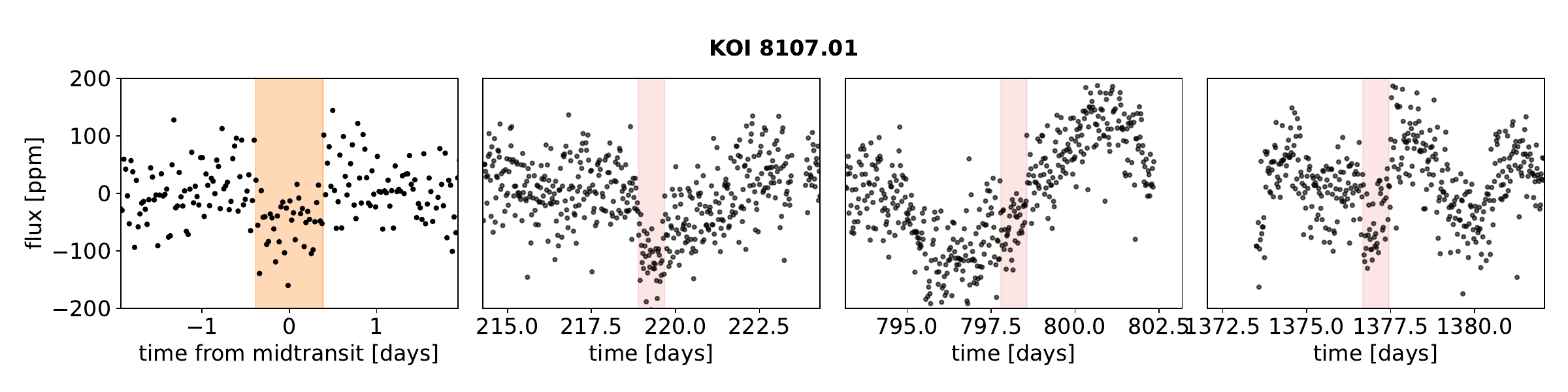}{
Light curve around the transits of the exoplanet candidate KOI 8107.01 with a Robovetter disposition score of $38\%$ and reliability of $62 \%$.
If it was confirmed, it would be a planet in the conservative habitable zone with radius $R = 1.2 R_{\oplus}$.
Shaded area is the transit region. The left-most panel is the folded light curve, which seems like a genuine exoplanet signature. The next three panels are individual transits of the candidate. The first and the third transits are apparently discontinuities in the light curve, which clearly reveals that KOI 8107.01 is a false alarm.
}

\DoFigure{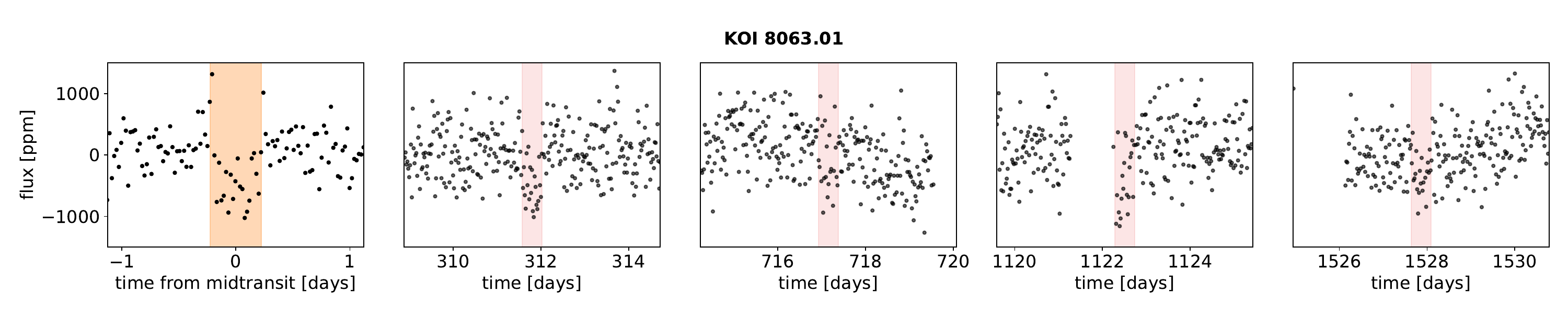}{
Light curve around the transits of the exoplanet candidate KOI 8063.01 with a Robovetter disposition score of $47\%$ and reliability of $80 \%$.
If it was confirmed, it would be a planet in the conservative habitable zone with radius $R = 2.1 R_{\oplus}$.
Shaded area is the transit region. The left-most panel is the folded light curve, which seems like a genuine exoplanet signature. The next four panels are individual transits of the candidate. The first and the third transits are apparently discontinuities in the light curve, which clearly reveals that KOI 8063.01 is a false alarm.
}

\DoFigure{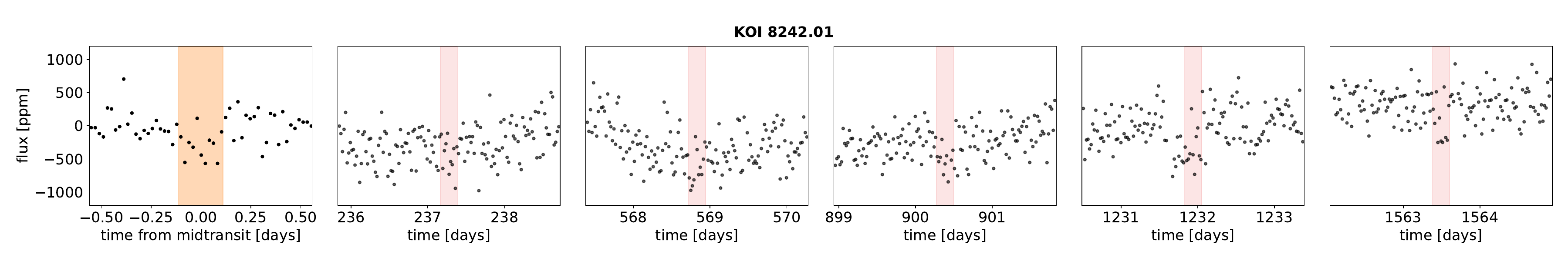}{
Light curve around the transits of the exoplanet candidate KOI 8242.01 with Robovetter disposition score of $18\%$ and reliability of $54 \%$.
If it was confirmed, it would be a planet in the conservative habitable zone with radius $R = 1.5 R_{\oplus}$.
Shaded area is the transit region. The left-most panel is the folded light curve, which seems like a genuine exoplanet signature. The next five panels are individual transits of the candidate. Fourth transit contributes a large amount to the candidate significance and seems like a localized defect, suggesting that KOI 8242.01 is a false alarm.
}

\DoFigure{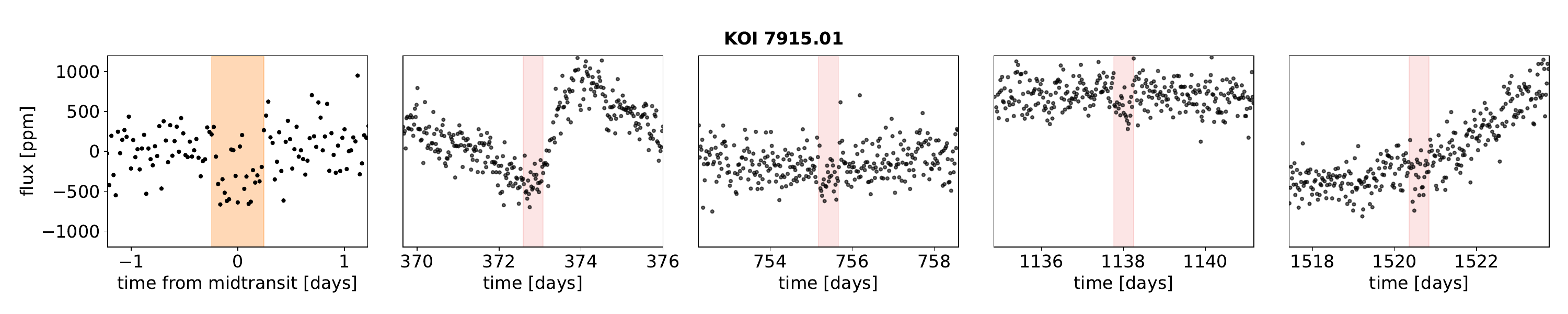}{
{%\color{red}
Light curve around the transits of the exoplanet candidate KOI 7915.01 with Robovetter disposition score of $32 \%$ and reliability of $46 \%$.
If it was confirmed, it would be a planet in the optimistic habitable zone with radius $R = 2.14 R_{\oplus}$.
Shaded area is the transit region. The left-most panel is the folded light curve, which could be a genuine exoplanet signature. The next four panels are individual transits of the candidate. 
%Fourth transit contributes a large amount to the candidate significance and seems like a localized defect, suggesting that KOI 8242.01 is a false alarm.
}}

As a first application of the newly developed pipeline we reanalyze the established planet candidates from \citet{christiansen_nasa_2025}. Out of total 3098 stars we eliminate known binary stars that appear in the Kepler eclipsing binary catalog \citep{prsa_kepler_2011} and the stars for which Gaia data is not available, leaving us with 2974 stars. 
{%\color{red}
If a star contains multiple planet candidates we remove all but the lowest SNR candidate, thus leaving us with one candidate per star.} We then analyze all stars without inputting the information about that candidate. Our pipeline independently finds all the planets with the disposition CONFIRMED. Large majority of them also pass the additional tests. There a some cases that are flagged by the transit significance consistency test. 
{%\color{red} 
We checked several of those cases and always found that the flag was caused by the planet having significant TTVs, which are not accounted for in the transit significance test.} 

On the other hand, our pipeline does not identify all \textit{unconfirmed} Kepler candidates as TCEs. Some of them do not even make it to the detection stage, because the region around some of their transits was identified as a localized defect and was masked out in Section \ref{sec: individual}. One such example is KOI 7894.01 from Figure \ref{fig: KOI 7894}, other examples are shown in Figures \ref{fig: KOI8107}, \ref{fig: KOI8063} and \ref{fig: KOI8242}.
Figure \ref{fig: candidates} shows the histogram of Kepler candidates in the radius-SNR plane. It demonstrates that while our pipeline agrees with the Kepler pipeline \citep{Jenkins2020, thompson_planetary_2018} at small periods or high SNR, it does not independently identify a majority of candidates at long periods with low SNR.

Robovetter's classification as a candidate or as a false alarm is binary but it comes with a score that measures the fraction of randomness realizations where the candidate disposition was obtained. 
{%\color{red} 
Candidates with high scores typically have high reliability, though candidates with low scores often also have high reliability \citep{bryson_reliability_2020}.} In Figure \ref{fig:robovetter} we show the probability that a given planet candidate is independently identified by our pipeline or considered a Kepler confirmed planet, given its Robovetter score. It shows that candidates with a high score (> 0.9) are indeed often confirmed, but this is not a strict requirement. On the other hand, candidates of any score may fail to be triggered by our pipeline, showing that our analysis uses the information that is not captured in the Robovetter score.

{%\color{red}
Table \ref{table} provides further information about the candidates that were not recovered by our pipeline. Its columns are divided in three groups. The first group lists general properties of the candidate, its 
Kepler Object of Interest (KOI) name and its period. 
The second group lists the results of our pipeline run, 
SNR of the candidate, SNR if we turn off the localized defect removal, false alarm probability (FAP), as determined by 50 runs of the null signal template (NST; \citet{robnik_reassessment_2025}) and booleans indicating if the candidate passed our false alarm vetting tests with and without the localized defects removal.
The third group consists of flags that can be used to judge the candidate. There is a flag indicating that a candidate has a low significance as determined by the Kepler pipeline (defined here as having SNR below 8.5 or Robovetter disposition score below 0.6) and a flag indicating if the candidate has a low significance as determined by our pipeline (defined here as having FAP above 0.5 or having failed the FA vetting tests). ``Defect'' flag shows if the candidate is likely caused by localized defects. It is set to True if either a significant drop in significance (change in SNR by more than 1) or vetting tests fail are observed after we remove the localized defects. Finally we also flag the candidates that have known TTVs from \citep{holczer_transit_2016}, as bona-fide planets with TTVs may fail our FA tests.
}
    
{%\color{red}
Of particular interest are potentially rocky planet candidates located in or close to the habitable zone (HZ). We use the list compiled by \citet{bryson_occurrence_2021}. The candidates on this list are drawn from 68,885 Kepler target stars that satisfy the conditions defined in \citet{bryson_occurrence_2021}, namely the availability of Gaia data, sufficient coverage of the Kepler time series, main-sequence classification, and effective temperatures in the range 3900–6300 K. Further, only candidates with radiaa in the range 0.5–2.5 $R_{\oplus}$ and with an instellation flux between 0.2 and 2.2 $I_{\oplus}$ are considered.} Finally, we here exclude candidates with VESPA astrophysical false-positive probabilities \citep{morton_vespa_2015, morton_false_2016} greater than 10\%, leaving a final sample of 44 candidates.

\begin{table*}
    \centering
    \caption{
    {%\color{red}
    Known Kepler candidates that were not independently recovered by our pipeline. This is a sample of the full table which is available online in a machine readable form. 
    The table provides information that can be used to determine the nature of these planet candidates. Column ``Defects'' denotes candidates that are likely caused by localized defects in the data. Column ``Low significance'' marks the canidates with a low statistical significance against false alarms.
    In this sample, the planet candidates from \ref{fig: hz} are shown. KOI 7894.01 and 8063.01 are identified as localized defects, because their SNR drops when the defects are eliminated. The other candidates are considered likely false alarms, because their FAP is larger than 50\%. Visual inspection of Figures \ref{fig: KOI8107} and \ref{fig: KOI8242} shows that KOI 8107.01 and KOI 8242.01 are likely caused by localized defects, despite not being identified as such in the table. This is because it is hard to automatically clearly distinguish these localized defects from planet transits.
    %The only exception is KOI 8107.01 for which we do not formally report the FAP
    }
    } \label{table}
    \csvreader[head to column names,
    tabular = ll|ccccc|cccc,
    table head = { 
    \bfseries KOI & 
    \makecell[t]{\bfseries Period \\ $[\mathrm{days}]$ } &
    \bfseries SNR & 
    \makecell[t]{\bfseries SNR \\ (defects)} & 
    \bfseries FAP & 
    \bfseries Vetting& 
    \makecell[t]{\bfseries Vetting \\ (defects)} & 
    \makecell[t]{\bfseries Low significance \\ (Kepler)}& 
    \makecell[t]{\bfseries Low significance \\ (our pipeline)}&
    \bfseries Defect & 
    \bfseries TTVs\\ 
    % & & [days] &  & & & (no defects removal) & (Kepler) & (our pipeline) & & &
    \midrule}, % column names
    table foot = \bottomrule
    ]{unrecovered_candidates_paper.csv}{}
    {\KOI & \period & \SNR & \SNRWithDefects & \FAP & \vettingTests & \vettingTestsDefects & \flagLowSignificanceKepler & \flagLowSignificance & \flagDefect & \flagTtv}
\end{table*}

Figure \ref{fig: hz} shows these candidates as a function of their radius, instellation flux and the host star effective temperature. Our pipeline eliminates 5 of those candidates. Three of these lie in the conservative habitable zone of Sun-like stars and have radius $1.2 R_{\oplus}$, $1.5 R_{\oplus}$ and $2.2 R_{\oplus}$. 
The remaining two lie in the optimistic habitable zone. 
Reliability \citep{bryson_probabilistic_2020} of these candidates is in the range from 46\% to 89\%, so they contribute significantly to the estimates for the occurrence of Earth-like planets in the habitable zone, i.e., $\eta_{\oplus}$. Their reliability is comparable to the reliability of confirmed planets in this part of the part parameter space, for example Kepler 452-b has a reliability of 68\%. Some of these candidates also achieve high Robovetter scores, for example KOI 7894.01 has a score of $84 \%$ which is higher than $77\%$ for Kepler 452-b. 
% Exominer does not prove to be useful for discarding those candidates either, as it assigns a low score to all of them (KOI 8063.01 has a score of $11 \%$ the other have scores below $1\%$) but so it does to the confirmed planets, such as Kepler 452-b, that gets a sub-percent score. 
Given the small total number of candidates in this parameter range, our finding that these candidates are false alarms could have a significant impact on $\eta_{\oplus}$ estimates.
{%\color{red}
Figures \ref{fig: KOI8107}, \ref{fig: KOI8063}, \ref{fig: KOI8242} and \ref{fig: KOI7915}, \ref{fig: KOI 7894} show the folded light curve and the individual transits for the three candidates in the conservative habitable zone and the two  candidates in the optimistic habitable zone respectively. Although the folded light curve seems to have a transit shape, the individual transits reveal that KOI 7894.01, 8107.01, 8063.01 and 8242.01 are likely caused by a combination of localized defects. KOI 7915.01 is not clearly caused by localized effects, but is nonetheless attributed a low statistical significance against false alarms with FAP of around 89\%, see Table \ref{table}.}

% We found 452-b and the other three confirmed planets that had $P_{planet} < 0.5$ in \citet{ivashtenko_independent_2025}, should we comment on this? But then we probably need to add NST?

\section{Discussion}

We have introduced a pipeline for detection of exoplanet transits in the Kepler data, which enhances the sensitivity to the planets at the detection limit compared to the Kepler pipeline. 
Our pipeline improves the removal of localized defects prior to the planet search, improves vetting at the level of individual transits and introduces a Bayes factor test statistic and an algorithm for extracting multiple candidates from a single detection run.

One of the most significant factors in improving the sensitivity of our pipeline is the treatment of the localized defects which can combine to resemble a planet transit. 
The strength of our localized drops removal algorithm is that it is sensitive to the specific shape of the individual transit events through the false alarm template bank that we have built. On the other hand, this is also a limitation, as there might be localized defects which are not included in our template bank, and could then more easily evade identification.

Therefore we have also applied the statistical significance test, which is complementary in this regard, i.e., it only examines the transit significance with respect to the planet template. A limitation of this test is that potential TTVs might also trigger a failed test, which we observed for some confirmed planets. If one is interested in detecting planets with TTVs at the detection limit \citep{leleu_alleviating_2021, carter_quasiperiodic_2013}, this test can easily me modified.

We have shown that all of these innovations improve the detection sensitivity. The reanalysis of the known planet candidates revealed that a large proportion of the known candidates with low SNR and long period are likely false alarms. In particular, some Earth-like planet candidates in the habitable zone are found to be likely false alarms, as is also confirmed by examining their individual transits by eye. Given the small number of candidates in this part of parameter space our finding could have a significant impact on their occurrence rate estimates, $\eta_{\oplus}$. This analysis is left to future work.

\section*{Acknowledgements}

This material is based upon work supported in part by the Heising-Simons Foundation grant 2021-3282. 
Kepler was competitively selected as the 10th Discovery mission and was funded by NASA's Science Mission Directorate. The authors acknowledge the eﬀorts of the Kepler Mission team in obtaining the light curve data and data validation products used in this publication. These data were generated by the Kepler Mission science pipeline through the eﬀorts of the Kepler Science Operations Center and Science Oﬃce. 
%The Kepler light curves are available at the Mikulski Archive for Space Telescopes, and the Data Validation products are available at the NASA Exoplanet Science Institute.

\section*{Data Availability}
The data underlying this article are available in NASA Exoplanet Archive, at  \url{https://exoplanetarchive.ipac.caltech.edu/bulk_data_download/}.

\bibliographystyle{mnras}
\bibliography{references,references2}

\appendix

\section{Template Bank} \label{appendix: bank}

Our goal here is to determine the optimal spacing of templates in the template bank, such that they ensure a predefined SNR loss tolerance at a minimal number of templates. In low dimensional parameter spaces, geometric lattice placement of the templates in optimal, whereas stochastic placement is typically used in high dimensions. In the context of exoplanets, both have been applied, \citet{jenkins_matched_1996} used the lattice, whereas \citet{ivashtenko_independent_2025} used the lattice for the period and the phase and stochastic placement for the other planet parameters. We argue that the transit duration is the only other parameter relevant for the planet detection \citep{robnik_matched_2021}, so we will adopt the lattice approach.
If a template with duration $\tau$ is used to compute the SNR of the signal $\tau'$ we get a fractional SNR loss of
\begin{equation}
    r(\tau, \tau') = 1 - \braket{S_{\tau}}{S_{\tau'}} / \braket{S_{\tau'}}{S_{\tau'}} \approx \frac{1}{2}  \vert \partial_{\tau} \boldsymbol{s}_{\tau} \vert^2 (\tau - \tau')^2,
\end{equation}
where the final step is obtained by Taylor expansion in $\tau - \tau'$, which is known as the Fisher analysis in this context.
Planet template depends on the transit duration only through the combination $t/\tau$, let's denote this by
\begin{equation}
    S_{\tau}(t) = \mathcal{S}(x = t / \tau).
\end{equation}
Assuming flat power spectrum and taking the derivative then shows 
\begin{equation}
     \vert \partial_{\tau} \boldsymbol{s}_{\tau} \vert = \frac{1}{\tau} \bigg( \frac{(\int x \mathcal{S}'(x) + \frac{1}{2} \mathcal{S}(x))^2 dx}{\int \mathcal{S}(x)^2 dx}\bigg)^{1/2} = \frac{C}{\tau},
\end{equation}
where $C$ is a $\tau$-independent constant.
We want to cover a domain $[\tau_{\mathrm{min}}, \tau_{\mathrm{max}}]$ by choosing a grid $\{ \tau_m \}_{m = 1}^M$. Our goal is to ensure the covarage, $r(\tau , \tau_m) < r_{\mathrm{max}}$ for all $\tau_{\mathrm{min}} \leq \tau \leq \tau_{\mathrm{max}}$ and some predefined tolerance $r_{\mathrm{max}}$, say $r_{\mathrm{max}} = 0.05$. The range of the domain that a single template covers is then $\Delta \tau$, such that
\begin{equation}
    \frac{1}{2} \bigg( \frac{\Delta \tau}{2}\bigg)^2 \, \frac{C^2}{\tau^2} = r_{\mathrm{max}}
\end{equation}
and so we may write for the number density of the templates:
\begin{equation}
    \frac{d n}{d \tau} = \frac{C}{\tau \sqrt{8 r_{\mathrm{max}}}} \propto 1 /\tau
\end{equation}
or uniform in $\log \tau$. 
An analogous argument can be made for templates with different periods.
The conclusion is therefore that we should have templates equally spaced in log in both the period and the transit duration, consistent with \citet{jenkins_matched_1996}. We determine the density of templates numerically, and find that 
70,000 periods per log in period and 20 templates per log in transit duration ensures SNR loss below 10\%. Interestingly, this is precisely the prescription used in \citet{ivashtenko_independent_2025}, where $T / \Delta \approx 1400 \,\mathrm{days} \, / \, 30 \, \mathrm{min} = 70,000$ is used for periods.

% The total number of templates that we need 
% \begin{equation}
%     n = \frac{C}{\sqrt{8 R}} \log \frac{\tau_{\mathrm{max}}}{\tau_{\mathrm{min}}} \approx 25
% \end{equation}
% if for example $\tau_{\mathrm{min}} = 0.1 \tau_{K}(P = 2 days) $ and $\tau_{\mathrm{max}} = 2 \tau_K(P = 400 days)$.

\section{Transit duration prior} \label{sec: tau prior}
We here derive the prior distribution of the transit duration. The strategy is to derive the transit duration as a function of planet's period, orbit orientation, eccentricity and stellar density (see Appendix \ref{sec: tau func}) and then marginalize over those parameters, given their prior distribution:
\begin{itemize}
    \item Isotropic \textbf{orbit orientation}.
    \item \textbf{Eccentricity} prior as fitted to the Kepler data in \citet{kipping_bayesian_2014}, i.e. a beta distribution with parameters $\alpha = 0.867$ and $\beta = 3.03$.
    \item Normal distribution for the \textbf{stellar density} $\rho_*$ with mean and variance taken from \citet{berger_gaia-kepler_2020}. 
\end{itemize}
The marginalization is performed by Monte Carlo integration, i.e. we draw the orbit orientation parameters, eccentricity and stellar density from their priors and calculate the transit duration for each of these samples, thus obtaining samples from the transit duration prior. We then calculate a histogram and fit it by a spline to obtain a smooth density.

% The only component of this prior that is specific to each planet candidate is the period of the planet and the stellar density. We will show that the mean of the stellar density distribution and the period only shift the transit duration prior, but the overall shape depends only on the relative error on $\rho_*$. It therefore suffices to precompute the transit duration prior on a fine grid of these relative errors and interpolate to save the computation at the pipeline run time.

\section{Transit duration as a function of orbital parameters} \label{sec: tau func}

Transit duration is a known function of orbital parameters, see e.g. \citet{kane_exoplanet_2012}, which we rederive here for completeness. 
{%\color{red}
Throughout we assume the planet's radius is small compared to the stellar radius.} The orientation of the planet's orbit can be parametrized by the Keplerian orbital elements, such that the direction of the normal to the planet's orbit is parametrized by inclination $i$ and the longitude of the ascending node $\Omega$ and the location of the periapsis is parametrized by the argument of the periapsis $\omega$. The argument of periapsis $\omega$ is measured relative to the ascending node. Transiting planets correspond to $\cos i < R_* / r$, where $r$ is the star-planet distance at the transit time and $R_*$ is the radius of the star. 

We start by deriving a transit duration for a circular ($e = 0$) orbit and then generalize to arbitrary orbits. 

\subsection{Orbit inclination}
For the inclined orbit, the shadow of the planet will not cross the star diametrically, but will follow a parallel line which is a distance $r \, \cos i$ away from the center, where $r$ is the distance between the planet and the star. Transit duration is proportional to the length of the line that is traced by the planet's shadow when it crosses the stellar surface, which is $\sqrt{R_*^2 - (r \cos i)^2}$. We fix the constant of proportionality by considering the aligned orbit and get:
\begin{equation}
    \tau(z) = \tau_K \sqrt{1 - b^2},
\end{equation}
where we have introduced the impact parameter $b(i) = r\, \cos i \,/\, R_*$ and $\tau_K$ is the transit duration of the perfectly aligned orbit:
\begin{equation}
    \tau_K = \bigg( \frac{3}{\pi^2 G \rho_*} P \bigg)^{1/3}.
\end{equation}
$b$ is uniformly distributed on the interval [-1, 1] because the orbit orientation prior is isotropic. We will have to pay extra care if the eccentricity is non-zero, as $r$ then additionally depends on $\omega$.

\subsection{Eccentricity}
For the eccentric orbit, the planet's speed is changing with the location on the orbit. Time of transit is inversely proportional to the velocity in the direction perpendicular to the eye of sight:
\begin{equation} \label{eq: svt}
    \tau = \frac{2 R_*}{v_{\perp}}.
\end{equation}
$v_{\perp}$ is determined by the angular momentum conservation:
\begin{equation}\label{eq: L conservation}
    L = r \, v_{\perp},
\end{equation}
where $r$ is the distance between the star and the planet and $L$ is planet's angular momentum divided by its mass. $r$ is given by a polar form of the ellipse
\begin{equation}
    r(\omega) = \frac{a \, (1-e^2)}{1 + e \sin \omega}.
\end{equation}
$a$ and $e$ are the semi major axis and the eccentricity of the ellipse respectively. By integrating the Equation \eqref{eq: L conservation} over one period of the revolution we can relate the angular momentum to the area of the ellipse:
\begin{equation} \label{eq: 2nd Kepler law}
    L P = 2 \pi a^2 \sqrt{1 - e^2}.
\end{equation}
By inserting Equations \eqref{eq: L conservation} and \eqref{eq: 2nd Kepler law} in the Equation \eqref{eq: svt} we get
\begin{equation}
    \tau(\omega) = \tau_K \frac{\sqrt{1 - e^2}}{1 + e \sin \omega}.
\end{equation}
$\omega$ is uniformly distributed on the interval [0, 2$\pi$].
By combining both eccentricity and the orbit inclination, we get:
\begin{equation} \label{eq: tau function}
    \tau(b, \omega) = \tau_K \sqrt{1 - b^2}\, \frac{\sqrt{1 - e^2}}{1 + e \sin \omega}.
\end{equation}
Transit duration is independent of $\Omega$. 
%It scales linearly with $\tau_K$, so the $P^{1/3}$ and $\rho_{*}^{-1/3}$ are only multiplicative factors that do not interact with the other parameters. 

One has to be careful in the case of non-zero eccentricity, as $r$ then additionally depends on $\omega$. $b$ can still be considered an independent, uniformly distributed variable if a non-uniform prior is adopted on $\omega$, such that $p(\omega) \propto 1/r(\omega)$. In Monte Carlo integration this will be achieved by importance sampling, that is, $\omega$ will be drawn from the uniform distribution, but the samples will be assigned weights $1/r(\omega)$. Intuitively, the smaller the radius, the larger range of inclinations gives rise to a transiting planet.

\section{Planet radius calculation} \label{sec: A2r}
{%\color{red}
Transits provide a lot of information about the planet's radius through the amplitude of the transit. The relative reduction in host star's brightness at the center of transit, $A_{\mathrm{host}}$ is calculated as the integral of stellar surface brightness over the planet's silhouette, relative to the total stellar flux:
\begin{align} \label{eq: A2R}
    A_{\mathrm{host}} &= \frac{\int_{(x - b R_*)^2 + y^2 < R^2} U(\sqrt{x^2 + y^2}) dx dy}{\int_{x^2 + y^2 < R_*^2} U(\sqrt{x^2 + y^2}) dx dy} \\ \nonumber
    &\approx \frac{U(b)}{1 - u_1/3 - u_2/6} \,\frac{R^2}{R_*^2}.
\end{align}
Here $b$ is the impact parameter of the transit, i.e. the closest projected distance between the planet and the center of the star in units of stellar radius $R_*$. $R$ is planet's radius. $U$ is stellar surface brightness
\begin{equation}
    U(b) = 1 - u_1 B(b) - u_2 B(b)^2, \quad B(b) = 1 -\sqrt{1 - b^2},
\end{equation}
as modeled by quadratic limb darkening law \citep{kopal_classification_1995} with limb darkening parameters $u_1$ and $u_2$. The amplitude of the transit $A$ is given by $A_{\mathrm{host}}$, but potentially blended by the other stars in the same or the neighboring pixels:
\begin{equation}
    A = A_{\mathrm{host}} \frac{F_{\mathrm{host}}}{F_{\mathrm{host}} + F_{\mathrm{other}}}.
\end{equation}
Here, $F_{\mathrm{host}}$ is the flux of the host star when not obstructed by the planet and 
$F_{\mathrm{other}}$ is flux that the other stars contribute to the target pixel. Flux of the stars which are not located in the target pixel is diluted by the point spread function of the Kepler telescope.
}

To summarize, the parameters that determine the radius are: amplitude, stellar radius, blend factor, limb darkening parameters and the impact parameter.
To generate samples from the marginal $R$ distribution, one can first generate samples for those parameters, given the observed light curve, and then calculate $R$ for each of those samples. The final $R$ measurement and its uncertainty is then summarized by the mean and the asymmetric errors.

The amplitude posterior given the data is Gaussian with mean and variance determined by the matched filter \citep{robnik_matched_2021}. Stellar radius posterior is taken from \citet{berger_gaia-kepler_2020}, where the mean and asymmetric errors are reported. Limb darkening parameters are calculated from the properties of the star: effective surface temperature, surface gravitational acceleration and metalicity \citep{claret_gravity_2011}. To generate samples for the impact parameter we use a procedure similar to appendix \ref{sec: tau prior}, except that now we reverse the roles of the transit duration and the impact parameter: we take the posterior distribution for the transit duration and marginalize over it with Monte Carlo integration to obtain samples from the marginal distribution of the impact parameter. 
{%\color{red} 
To compute the flux of blending stars we follow TRICERATOPS \citep{giacalone_vetting_2021} and query the Gaia catalog for nearby stars. To those we add a random draw from the background population of stars not detected by Gaia. This undetected population is simulated by the TRILEGAL simulation \citep{girardi_star_2005}. $F_{\mathrm{other}}$ is then calculated by adding the fluxes of all of those stars, weighted by the point spread function.}

\begin{figure}
    \centering
    \includegraphics[width=\linewidth]{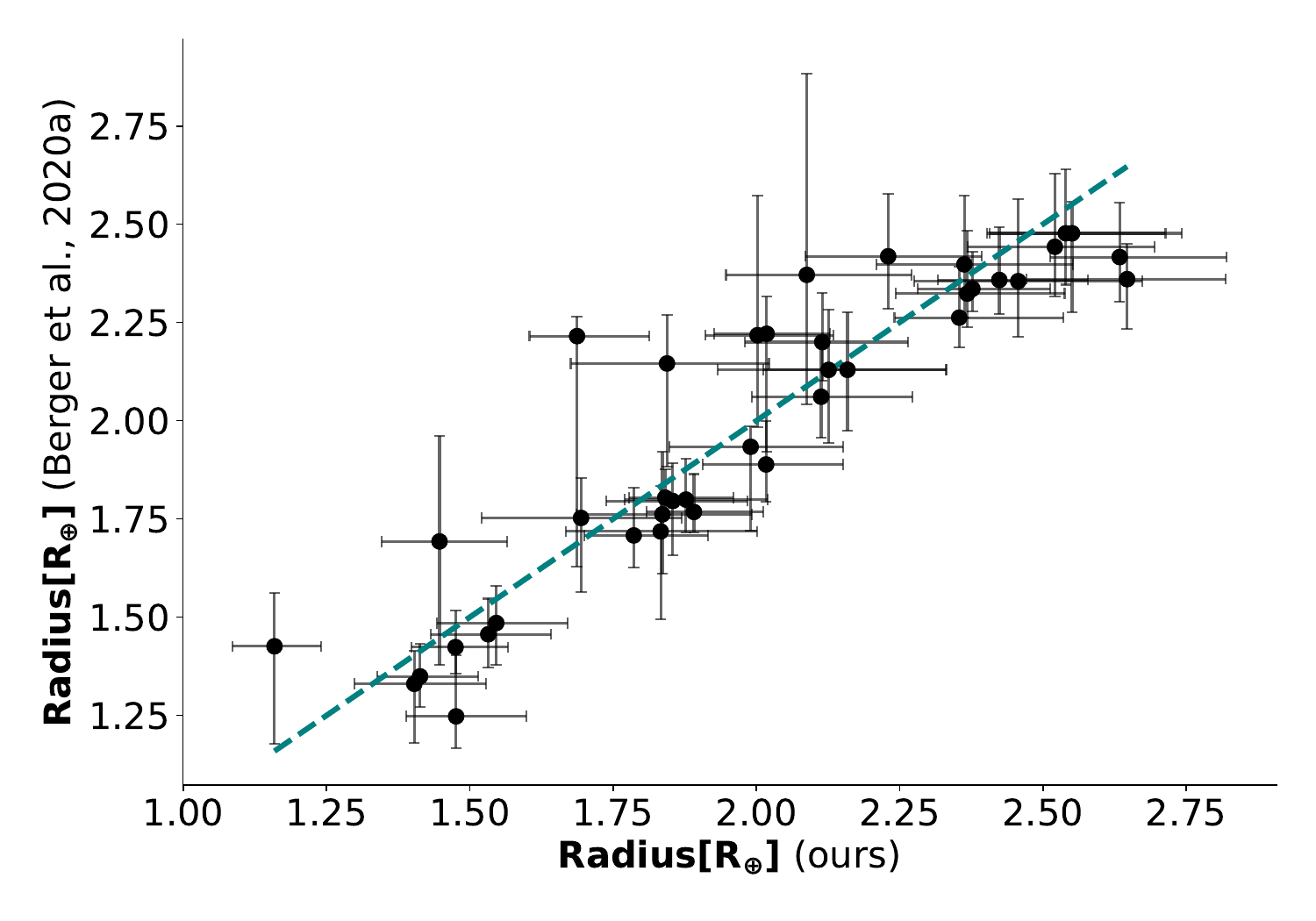}
    \caption{Radius of the planet candidates from Figure \ref{fig: hz}, i.e., the Earth-like and super Earth planets in or close to the habitable zone. Radius calculated by our pipeline, as described in Appendix \ref{sec: A2r}, is compared to the literature values from \citet{berger_gaia-kepler_2020}, showing that the two are consistent.}
    \label{fig: radius}
\end{figure}

In Figure \ref{fig: radius} we compare our results with values from \citet{berger_gaia-kepler_2020}, showing that the two methods are consistent.

\section{Non-stationary matched filter} \label{sec: nonstat}

Noise in the Kepler data is not strictly stationary, due to the time-varying stellar variability and rolling bands which appear on some channels, but not on the others. 
In this appendix we develop a non-stationary matched filter (NSMF) to take this into account and test it on injections in the Kepler data. Ultimately, we show that it does not bring any benefits compared to the stationary matched filter, but does no harm either. We therefore do not use it in the main pipeline. 

\subsection{Method}
NSMF allows for a slowly time varying power spectrum. The slow time-dependence ensures that the power spectrum is locally approximately stationary. The $SNR$ of a transit event located at time $t$ can then be calculated as
\begin{equation}
    SNR(t \vert x, S) = \frac{\braket{x}{S}_t}{\braket{S}{S}_t^{1/2}},
\end{equation}
where $x$ are the data, $S$ is the planet transit template and we have defined the scalar product
\begin{equation}
    \braket{x}{y}_t = \sum_{k} \frac{\mathcal{F}^*(x)_{k, t} \, \mathcal{F}(y)_{k, t}}{\mathcal{P}_{k, t}}
\end{equation}
analogously to Equation \eqref{eq: scalar product}. We have used the Short-time Fourier transform (STFT),
\begin{equation}
    \mathcal{F}^*(x)_{k, t} = \frac{1}{\sqrt{N_w}} \sum_{t'} w_t(t') x(t') e^{2 \pi i k t' / N}.
\end{equation}
The window function $w_t(t')$ is inserted in the usual Fourier transform to isolate the short span of the data around $t$. We here adopt the Hanning window $w_t(t') = 1 + \cos \pi (t' - t) / t_w$ with the half-width of $t_w = 25$ days. The normalization constant $N_w = \sum_{t'} w(t')^2$ is the effective number of data points in the window. Naively one could estimate the instantaneous power spectrum $\mathcal{P}_{k, t}$ as
\begin{equation}
    \mathcal{P}^{(naive)}_{k, t} = \vert \mathcal{F}(x)_{k, t}\vert^2,
\end{equation}
but this estimate would be biased by the presence of the potential planet transit at $t$. We instead average the power spectrum over two neighboring windows, which are placed $t_w$ away from the transit, so that the transit does not contaminate the estimate:
\begin{equation}
    \mathcal{P}_{k, t} = \frac{1}{2}\bigg(\vert \mathcal{F}(x)_{k, t - t_w}\vert^2 + \vert \mathcal{F}(x)_{k, t + t_w}\vert^2 \bigg).
\end{equation}
If one of these two windows has a large portion of missing data (> 50\%) we deem the power spectrum estimate there unreliable and do not include it in the average. If both windows have large portion of missing data, we use a stationary estimate for the power spectrum, rather than the instantaneous one.
Finally, we smooth the power spectrum by a moving average in the frequency domain, as described in Section \ref{sec: PSD estimation}.

Combining multiple transits of a planet located at times $\{ t_n \}$ results in a SNR:
\begin{equation}
    SNR(\{ t_n \} \vert x, S) = \frac{\sum_n \braket{x}{S}_{t_n}}{( \sum_n \braket{S}{S}_{t_n} )^{1/2}}.
\end{equation}
Note that the local power spectrum is taken as a weight in the matched filter, which can be useful if power spectrum changes significantly over time. In the rolling band example, the power spectrum would increase during the noisy epochs and the transits there would be down-weighted relative to the other transits. 

NSMF approach can be compared to the wavelet filter \citep{jenkins_tests_2002, Jenkins2020}, where power spectrum is also time dependent. 
{%\color{red}
The main difference is that in the wavelet approach the time and frequency bands are specified in advance, for example by Daubechies’ 12-tap wavelets \citep{daubechies_orthonormal_1988, Jenkins2020}\footnote{Dyadic wavelets correspond to a filter bank with a constant ``quality factor'' in that the ratio of the bandwidth to the center frequency is constant.}. In NSMF, time band width is independent of the frequency, whereas frequency band width depends on the frequency in an adaptive manner. }

\subsection{Results}
\begin{figure}
    \centering
    \includegraphics[width=\linewidth]{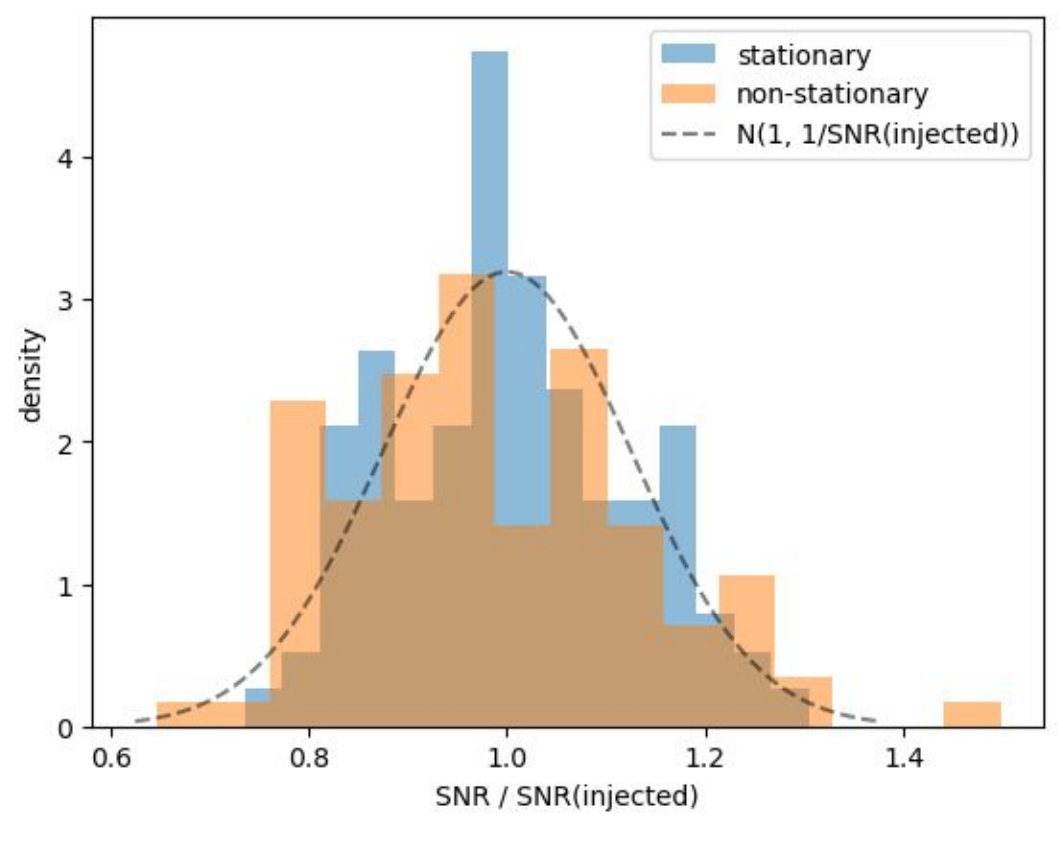}
    \caption{SNR recovered from simulations with stationary stellar variability noise, relative to the injected SNR. Both stationary and non-stationary matched filters perform equally well, are unbiased and match the Gaussian distribution prediction for the stationary matched filter.}
    \label{fig: stat test}
\end{figure}

\begin{figure}
    \centering
    \includegraphics[width=\linewidth]{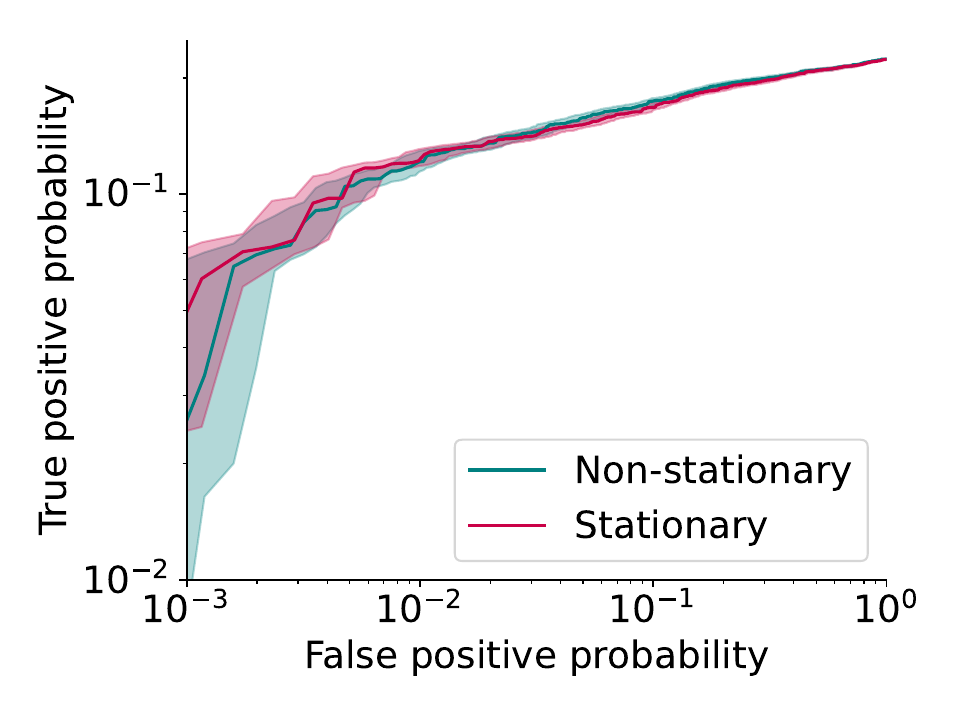}
    \caption{ROC curve comparing the nonstationary and stationary matched filter, analogously to Figure \ref{fig:roc}. Both perform similarly.}
    \label{fig: roc nonstat}
\end{figure}

\paragraph*{Stationary simulations} As a first test of the non-stationary matched filter we apply it to the stationary simulations with correlated Gaussian noise, where stationary matched filter is known to be optimal. 
Our simulated light-curves consist of $1400$ days of flux measurements which are equally spaced in time with half an hour increments. The gaps in the data match the Kepler-90 star light curve. Noise power spectrum is extracted from the Kepler 90 data \citep{robnik_matched_2021}. We then inject a planet with a period of $100$ days, $SNR = 8$ and the other parameters randomly drawn from the prior. We simulate 50 light curves. Figure \ref{fig: stat test} shows the $SNR$ retrieved by our pipeline using either stationary or non-stationary matched filters. Both perform equally well, demonstrating that the non-stationary filter is close to optimal, even for the stationary noise.

\paragraph*{Kepler data} We compare the matched filters on injections in the actual Kepler data, analogously to Section \ref{sec: ROC}. Figure \ref{fig: roc nonstat} compares their ROC curves, which are very similar, indicating that there is no benefit in using the non-stationary matched filter.

\section{Gaussianization transformation} \label{appendix: gaussianization}

Here we review the construction of the Gaussianization transformation $\Psi(x)$ from \citet{robnik_matched_2021} used to eliminate the impact of non-Gaussian outliers in the Kepler photometric data. The transformation is designed such that:

\begin{enumerate}
    \item If $x$ consists of white, uncorrelated non-Gaussian noise, then $\Psi(x)$ becomes a standard Gaussian.
    \item If $x$ is part of a correlated astrophysical signal (e.g., a planet transit), $\Psi(x) \approx x$, preserving the shape and amplitude of the signal.
\end{enumerate}

The white noise component of the data is modeled as a mixture distribution:
\begin{equation} \label{eq: q pdf}
    q(x_i) = (1 - a) \, N(x_i) + a \, \mathrm{NCT}(x_i),
\end{equation}
where $N$ is a probability density distribuition (PDF) of a zero-mean Gaussian distribution, $\mathrm{NCT}$ is a PDF of a non-central t-distribution capturing the heavy-tailed outlier population, and $a \ll 1$ is the outlier probability, typically of order $10^{-3}$ in Kepler data \citep{robnik_matched_2021}. These outliers are uncorrelated and differ statistically from the rest of the white noise.
The Gaussianization $\Psi$ is a local, non-linear function applied pointwise or quasi-pointwise to the data.

\textbf{1D Gaussianization: }
in the absence of correlated structures, the transformation $\Psi$ reduces to a purely pointwise mapping $\psi^{(1d)}$, which is defined by matching the cumulative distributions:
\begin{equation}
    \psi^{(1d)}(x_i) = \mathrm{CDF}_N^{-1} \circ \mathrm{CDF}_q(x_i),
\end{equation}
where $\mathrm{CDF}_N$ is the cumulative distribution function (CDF) of a standard normal distribution, and $\mathrm{CDF}_q$ is the CDF of the mixture distribution $q$ from Equation \eqref{eq: q pdf}.
This transformation maps the heavy-tailed noise distribution into a standard Gaussian, thus suppressing the impact of outliers.

\textbf{Preservation of Correlated Structures: }
To avoid affecting real astrophysical signals (such as planet transits), which are also non-Gaussian but \textit{correlated}, we modify the transformation to distinguish between isolated outliers and contiguous structures. This is achieved using a soft gating function $\mathscr{P}_i(x)$:
\begin{equation}
    \Psi_i(x) = \mathscr{P}_i \, x_i + (1 - \mathscr{P}_i) \, \psi^{(1d)}(x_i),
\end{equation}
where:
\begin{equation}
    \mathscr{P}_i(x) = 1 - P(x_{i-1} \text{ and } x_{i+1} \text{ not outliers} \mid x_{i-1}, x_{i+1}).
\end{equation}

Using Bayes' theorem and the mixture model $q(x_i)$, the conditional probability that a point is \emph{not} an outlier is:
\begin{equation}
    P(\text{not outlier} \mid x_i) = \frac{(1 - a) \, N(x_i)}{(1 - a) \, N(x_i) + a \, \mathrm{NCT}(x_i)}.
\end{equation}

Putting this together, we have:
\begin{align} \nonumber
    \mathscr{P}_i &= 1 - P(\text{i-1 not outlier} \vert x_{i-1}) \cdot P(\text{i+1 not outlier} \vert x_{i+1}) \\
    &= 1 - \prod_{j \in \{i-1, i+1\}} \frac{(1 - a) \, N(x_j)}{(1 - a) \, N(x_j) + a \, \mathrm{NCT}(x_j)}.
\end{align}

This scheme ensures that isolated large deviations (likely outliers) are Gaussianized, while contiguous ones (likely real transits) are preserved.

%\bsp	% typesetting comment
\label{lastpage}
\end{document}